\documentclass[12pt,a4paper]{article}
\usepackage[dvips]{graphicx}
\usepackage{xspace,epsfig}
\title{A three-dimensional calculation of atmospheric neutrinos}
\author{G.D.~Barr$^1$, T.K.~Gaisser$^2$, P.~Lipari$^3$, S.~Robbins$^1$, 
        T.~Stanev$^2$\\
\small\it $^1$ Department of Physics, University of Oxford,\\ 
\small\it      Denys Wilkinson Building, Keble Road, Oxford, UK, OX1 3RH\\
\small\it $^2$ Bartol Research Institute, University of Delaware,\\
\small\it      Newark, Delaware, USA 19716\\
\small\it $^3$ Dipartmento di Fisica, Universit\'{a} di Roma (La Sapienza)\\
\small\it      and INFN, Sezione di Roma, P.~A.~Moro 2, I-00185 Roma, Italy}
\begin{document}
\maketitle
\begin{abstract}
A Monte-Carlo calculation of the atmospheric neutrino
fluxes~\cite{BGS,AGLS} has been extended to take account of
the three-dimensional (3D) nature of the problem, including the
bending of secondary particles in the geomagnetic field.  Emphasis
has been placed on minimizing the approximations when introducing the
3D considerations.  In this paper, we describe the techniques used and
quantify the effects of the small approximations which remain. 
We compare 3D and 1D calculations using the same physics input in order
to evaluate the conditions under which the 3D calculation is required
and when the considerably simpler 1D calculation is adequate.  
We find that the 1D and 3D results are essentially identical
for $E_\nu>5$~GeV except for small effects in the azimuthal
distributions due to
bending of the secondary muon by the geomagnetic field
during their propagation in the atmosphere.
\end{abstract}

\section{Introduction}

The hypothesis that neutrino oscillations are observed in the fluxes of
muon-neutrinos produced from cosmic ray interactions in the upper
atmosphere~\cite{Kam,IMB} has held up well in analysis of high
statistics data from Super-Kamiokande~\cite{SuperK} and at other
experiments~\cite{Soudan-2,MACRO}.  What was once called an
`atmospheric neutrino anomaly' is now accepted as an established
demonstration of neutrino mass~\cite{KajTot}.  The atmospheric neutrino
oscillation result is obtained by comparing measured fluxes of
muon- and electron-neutrinos in underground detectors with computations based on
modeling hadronic interactions in the atmosphere over the surface of
the globe.  The angular dependence of the muon to electron
ratio and its energy dependence probe a range of 
nearly 5 orders of magnitude in L/E.  Deviations from the
expected behavior, in particular a deficit of muon neutrinos, 
point to oscillations in the $\nu_\mu\leftrightarrow\nu_\tau$ sector.

Until recently, only one-dimensional (1D) calculations of the
neutrino flux have been used to infer oscillation parameters
from the data.
In the 1D approximation, all interactions
and decay products follow the direction of the incident 
cosmic-ray particle that produced them.
This approximation simplified the problem
so that it could be tackled on computers of that era.  Even now, a
calculation which removes this limitation requires careful choice of
technique and considerable computer time to be successful.
     
It is known that the 1D approximation neglects a geometrical effect
which dramatically changes the predicted zenith angle distributions at
low energies~\cite{Lipari1} (see figure~\ref{fig:zenkam}).  However,
in oscillation studies, the neutrino spectrum is cut at low
energies by detector acceptance, smeared due to the experimental
determination of the neutrino direction and weighted by the
neutrino cross section which increases (approximately linearly) with neutrino
energy.  These effects combine to reduce the importance of a full
three-dimensional (3D) calculation to the point where its effect on
the extraction of oscillation parameters is expected to be
slight even though it
considerably complicates the calculation and data analysis procedure.
Nevertheless, in view of the importance of the result, it is essential
to use fully three-dimensional calculations for interpretation of
the data to infer the oscillation parameters.

In this paper we extend our original 1D calculation~\cite{BGS,AGLS} to
include a full 3D treatment of showers across the whole surface of the
globe.  Our goal has been to make a code that is
sufficiently fast to be able to investigate systematically the choices
used in making an accurate calculation without the 1D approximation.
We compare the 3D and
1D calculations in detail (along with various intermediate steps)
in order to display the origin of characteristic features of the 3D
calculation.  We also identify the situations in which the 1D
results adequately approximate those of the 3D calculation.  
In particular, we
will identify the neutrino energy above which, 1D calculations can
still be used.  

This paper deals with the technical aspects of
moving from a 1D to a 3D calculation.  In \S2 we 
describe the steps in our calculation
in the context of previous 3D calculations.  In \S3 we
present and discuss comparisons between 1D and 3D results
as a function of neutrino energy and direction (zenith and azimuth).
We also summarize some important technical aspects of a
3D calculation as well as the differences between 3D and 1D results in \S3.
Comparison of our calculated neutrino fluxes with others and
evaluation of the larger uncertainties caused
by different choices of hadronic models and differences
among measurements of the primary cosmic-ray spectrum will
be given in a later publication.

\section{Overview of calculation}

The neutrino flux is a
convolution of the primary cosmic ray flux with the neutrino yields 
from interactions of the cosmic rays in the atmosphere.
In general, the directional dependence of the flux at the detector is
obtained by generating showers with random positions over 
the globe and collecting the neutrinos that pass
through the detector.  See Ref.~\cite{GH} for a review.

As a consequence of the geomagnetic field, 
the primary cosmic-ray spectrum incident on the 
atmosphere depends on location.  In a 1D calculation, the
assumption is that all secondaries follow the direction of
the primary particle that initiated the cascade in
which they were produced.  In the 1D calculation, therefore,
the geomagnetic field can be accounted for simply by
evaluating the geomagnetic cutoff at each grid point on the
globe for the single direction that points toward the detector.
Moreover, the 1D calculation is extremely efficient because
only cascades pointed at the detector need be generated.
In reality, however, the secondaries deviate from 
the direction of the primaries.  In a 3D calculation, therefore,
one must sample incident particles from all directions at each
point on the globe.  The efficiency of a fully 3D calculation
then is of order $A/R_\oplus^2\sim 10^{-10}$, where $A$ is the
projected area of the detector and $R_\oplus = 6372$~km the radius
of the Earth.  Because the center of the geomagnetic dipole is
offset from the center of the Earth, and because the field is
not an exact dipole, any symmetry technique for making the calculation
manageable involves an approximation that is difficult to quantify.

Deviations of cascade particles from the direction of the primary
have two sources.  First is the transverse momentum characteristic of 
hadronic interactions ($\sim300$~MeV/c for pions) and the decay
processes such as $\pi^+ \rightarrow \nu_\mu\mu^+$ followed by 
$\mu^+ \rightarrow
e^+\nu_e\overline{\nu}_\mu$ in which the neutrinos are produced.
The scale of this deviation is set by pion production as
\begin{equation}
\label{deviation}
{\langle p_T\rangle\over E_\pi}\;\sim\;{300\;{\rm MeV}\over E_\pi}\;\sim\;
{0.1\over E_\nu\;{\rm GeV}}\;{\rm radians}.
\end{equation}
Characteristic 3D effects are therefore most important for neutrinos
with sub-GeV energies.

In addition, there is a second source of deviation from the direction
of the primary, which is the bending of muons in the geomagnetic field.
In this case, because gyro-radius and decay length have opposite dependence
on energy, the deviation is independent of energy and is typically
of order $3^\circ$.  The muon decay length is 
$\gamma\,c\,\tau_\mu\;\approx\;6.2\,{\rm km}\times E_\mu({\rm GeV})$
compared to typical production altitudes of 15 km, so only muons 
with several GeV and above begin to hit the ground before decay.  This deviation
from the 1D approximation therefore remains important up to
high energy, particularly for large zenith angles where higher 
energy muons decay before reaching the ground.

Bending of primary cosmic-rays before they interact, as well as
energy loss of muons and protons in the atmosphere must also be
accounted for.

\subsection{Survey of 3D neutrino flux calculations}

Wentz et.~al.~\cite{Wentz} give 
a comprehensive summary of calculations of the neutrino flux,
both 1D and 3D.
Here we comment on 3D calculations, noting technical assumptions
and approximations that have been made by the authors for
comparison with our approach.

Battistoni et al.~\cite{Bat1} made the first calculation showing 
the characteristic enhancement of low-energy neutrinos
near the horizontal.  The calculation was updated~\cite{Bat2}
with emphasis on use of the FLUKA interaction model~\cite{FLUKA}.
The calculation ignores the geomagnetic field for all tracking
within the atmosphere.  This approximation allows the shower to be
developed at an arbitrary position on the globe, then moved such that
one of the neutrinos hits the detector.  The cutoff energy (see below)
is checked after the location of the primary is fixed and the
event rejected if the primary is below the cutoff rigidity.
This procedure is efficient because each cascade has a high 
probability of generating a neutrino that is used.

Lipari~\cite{Lipari1, Lipari2} performed a 3D calculation in which the
particles were injected over the entire Earth's atmosphere.  The
detector is represented by a region of $1/5$ the surface of the Earth.
The paper emphasizes that not only bending of muons but also bending of
protons in the geomagnetic field is important.

Honda~et.\ al.~\cite{HKKM3} use a dipole magnetic field approximation
which allows them to invoke the symmetry in the geomagnetic longitude
to increase the collection efficiency at the detector.  Many details
of the consequences on the azimuthal differences introduced by the 3D
calculation and of the path length distributions are addressed in this
paper.

Tserkovnyak et al.~\cite{Waltham} do a full 3D simulation similar to
our own with an enlarged rectangular detector $10^\circ \times 40^\circ$ 
with the narrow direction aligned with magnetic north.  This corresponds
to an effective detector area of about 1\% of the surface area of the Earth.
While they use a large surface area, Tserkovnyak et al. point out that
it is important not to enlarge the vertical dimension of the detector.
Doing so tends to wash out the enhancement near the horizon. 

Wentz et al.~\cite{Wentz} also do a full
3D simulation using the CORSIKA simulation
package~\cite{CORSIKA}.  Calculation of neutrinos from below
is done by injecting primaries over the whole Earth and collecting
neutrinos that pass within a circle of radius $1000$~km of 
Super-Kamiokande.  Downward neutrinos were calculated from locally injected
primaries.

Liu et al.~\cite{Liu} inject particles over an injection sphere at 2000 km
above the surface of the Earth.  They then calculate the
neutrino fluxes averaged over all azimuth in three bins of
geomagnetic latitude.  For comparison with Super-K they use a
spherical section 15 degrees wide in latitude by 30 degrees
in longitude.

With one exception, all 3D calculations (including ours) start by injecting
particles near the top of the Earth's atmosphere.  They then use backtracking,
as described below, to check whether the chosen energy and direction
of a particle is on an allowed trajectory, rejecting those that are not.
Playskin~\cite{Plyaskin} has attempted a much more ambitious calculation.
He injects particles at $10\times R_\oplus$ and follows their trajectories
to see which ones interact in the Earth's atmosphere, presumably a tiny
fraction of the total.  The results of this calculation differ significantly
from others for reasons that have so far not been well-understood.

Favier et al.~\cite{Favier} use an injection sphere at 380 km and
calculate the flux in 3 zones of geomagnetic latitude. For comparison
with Super-K they average over all azimuth within a band of
geomagnetic latitude centered on the detector.  They check
by comparing to the flux limited to $\pm$30 degrees in longitude
about the location of Super-K.  A significant aspect of
this paper is a comparison of the proton spectrum at the injection sphere
obtained with the backtracking method with that obtained by injecting
particles at $20\;R_\oplus$.  The two methods agree, including
reproduction of the ``second spectrum'' \cite{AMSsecond}, thus giving a nice
empirical confirmation of the backtracking method.

The diversity of techniques and results 
among the 3D calculations arises in part because of
the difficulty of the computational problem.  Detectors are small
compared to the size of the Earth, and 
there is no symmetry to the problem which can be
invoked without introducing some uncertainty.
The emphasis of the current study is first, to provide a fast
code which can be run in many different configurations to investigate
the importance of changes in the parameters and approximations used
and second,  to be accurate,
by which we mean that the
calculation should involve no approximations which affect the results by more than a
few percent.  

\subsection{Details of this calculation}

The calculation proceeds by running Monte-Carlo simulations of the
interactions of the primary cosmic rays with the atmosphere.  Separate
runs are performed at fixed primary energies, prearranged in
logarithmic steps in energy, 10 energies per decade from 1~GeV to
10~TeV.  The separate runs allow the details of the primary flux and
of the effects of the solar wind to be inserted at a later step when
the runs are combined.  In this way different primary spectra
and different solar epochs can be treated without re-running the full shower 
simulation. The statistics of each run is
determined by accumulating a fixed number of neutrinos.

The Earth is assumed to be a sphere of radius $R_\oplus=6372$~km.
Primaries are generated with random positions and angles on the
injection sphere which is at a radius of $r = R_\oplus+80$~km.  Each
primary which produces at least one neutrino in the detector is then
subjected to the cutoff calculation (see below) which takes account
of the effect of the Earth's magnetic field on the primary cosmic
rays.

The superposition approximation is used to treat primary nuclei,
i.e. the interaction of a projectile nucleus of mass $A$ and total 
energy $E$ is assumed to be
equivalent to the sum of $A$
individual nucleons interacting on the target
air nucleus individually, each with energy $E/A$. 
Three runs are generated at each primary
energy: with free primary protons; with bound primary protons; with
bound primary neutrons.  This is so that both bound protons and bound
neutrons can be propagated in the Earth's magnetic field as if they
were included in a nucleus with $A/Z = 2$.

The effect of the geomagnetic field on the interacting cosmic rays is
included by applying the cutoff using the back-tracing technique.  
Consider a point $A$ 
far from Earth where the
flux is assumed to be isotropic.  For each valid cosmic ray trajectory
from point $A$ to a point $B$ on the
injection sphere (radius $R_\oplus+80$~km), Liouville's theorem assures
that, since the phase space density at $A$ is isotropic, then the phase
space density at $B$, near the Earth is also locally isotropic around
the direction of the trajectory.

The technique is to select the particle on the injection sphere
isotropically and then backtrack it from B to A to see if it is indeed
on a valid trajectory -- i.e. that it projects back to a place far
from the Earth. A trajectory is considered valid when the particle
propagates to a distance of $30\;R_\oplus$ with a total path length
shorter than $300 R_\oplus$ without hitting the
Earth.  If this is true, then the neutrino is accepted.  If the
primary particle spirals back and either remains in the vicinity of
the Earth or hits the Earth itself, then its neutrino is discarded.

During the backtracking process, it is important to use an adjustable
step size depending on the local radius of curvature and the variation
of the Earth's magnetic field; otherwise the calculation takes a long
time or is inaccurate.  In the present calculation, the step size
adjusts itself by a factor of up to 100 along a particle trajectory.
The NASA parameterizations of the Earth's magnetic
field~\cite{Tsyganenko} have been used.

If a particle trajectory loops into the atmosphere and back out, then
one of the conditions for 
Liouville's theorem is no longer satisfied because the particle may
lose energy or interact.  In fact, interactions of protons
skimming the atmosphere produce albedo protons that can be
below the local geomagnetic cutoff.  This is the source
of the ``second spectrum'' measured in detail by AMS~\cite{AMSsecond}.
In a 1D calculation, this cannot happen 
by definition -- once the particle has started to interact with the
atmosphere, all its progeny follow in a straight line (towards the detector) and all
cosmic rays produced in the Earth's atmosphere are accounted for.  In
the 3D calculation we track all particles out to a sphere at radius
$R_\oplus+400$~km.  Those which loop back before
reaching this radius and interact are included, but
the small number of secondary particles which spiral
above this are lost.  The effect of this loss is small because the
second spectrum itself has a relatively low intensity~\cite{Zuccon}. 

Tracking within the atmosphere is carried out with a separate
algorithm; the center of the Earth is the origin of our global
coordinate system.  The injection point of each primary is defined in
this global coordinate system; then all points within a cascade are
tracked relative to that starting point so as not to lose accuracy by
combining large and small numbers.  Double precision is used
throughout.  Whenever a neutrino is generated, we return to the global
coordinate system to track the neutrino through the Earth.

Stepping is done for a fixed length $d{\ell}$ along the trajectory of
each particle ($d{\ell}=300$~m) except for kinetic energy below 200~MeV
where energy loss is large or altitudes below 10~km where the
atmospheric density is high (for which $d{\ell} = 30$~m).  At the end
of each track step, the local zenith angle is recalculated to take
account of curvature of the Earth.  From the altitude at the beginning and
end of each step, the local atmospheric density and hence $dx$ (in
g/cm$^2$) is computed.  For the present paper, the US standard
atmosphere model is used over the entire surface of the Earth to
obtain the variation of density with altitude.  This is a good average
of a more detailed model which includes seasonal and latitude
dependent effects which is currently being implemented.  The
calculation neglects the elevation of the land masses and treats the
surface of the Earth as being at sea level over the whole surface.
(Detectors, however, are assigned their actual altitude.)

Charged particles bend in the magnetic field as they are tracked
through the atmosphere.  An internal proper-time clock is maintained
for unstable particles to determine their decay point.  Likewise,
interacting particles are tracked until they have passed through the
appropriate thickness of atmosphere at which point they interact.  For
pions and kaons, decay and interaction are competing processes,
and the choice is made based on a comparison of the randomly
chosen interaction and decay lengths.

The decay and interaction generators are modular in this code, so they
can be plugged in and out in any combination.  For the investigations
in this paper, we use the interaction generator TARGET
version~2.1~\cite{Target21}.

The results we present in this paper are obtained with the primary
cosmic ray flux used by Agrawal et al~\cite{AGLS}.
Compared to the newer data sets this representation of
the primary spectrum has a lower proton flux then
was measured by AMS~\cite{AMS} and BESS~\cite{BESS} below 50 GeV, 
but higher then
the ones measured by other experiments~\cite{CAPRICE98}.  Another
feature of this primary spectrum is that the all-nucleon flux has a
flatter energy dependence at high energy
than fits including the AMS and BESS data
would give.  How the primary spectrum extrapolates to high
energy will be important for extension of the neutrino flux
into the TeV region.  

In the 3D calculation, particles are injected across the entire globe
and most neutrinos miss the detector and are not used to compute the
flux.  It is this reason why there is a huge step in complexity
between a 1D and a 3D computation.  To save computer time, a detector
which is many times larger than the real detector is used in the
simulation.  Tuning the shape and size of the detector is crucial for
a successful calculation.  Following the advice from
Ref.~\cite{Waltham}, we use a ``flat'' detector (i.e. a section from
a spherical shell with curvature given by the sum of the radius of the
Earth and the detector altitude).  The difficulty with any other choice
is that any 
significant enlargement in the vertical direction causes the detector
to either poke out the top of the atmosphere or be so deep that the
effects of the 3D geometry are not correctly represented.  The
question of how large the artificial detector in the calculation can
be made, without introducing excessive inaccuracies, is addressed later
in this paper.  We employ a circular shape with radius 500~km, centered
on the detector. Averaging of the neutrino flux over this distance
introduces less than 0.5\% bias in the results.

A further difficulty with using a truly flat detector is that the
cross sectional area changes as a function of the zenith angle of the
neutrino ($\theta_z$), and vanishes for horizontal directions.  This
effect can be solved in the 1D case by generating
primaries with a flat zenith angle distribution ($\theta_p$).  Since,
for a 1D calculation, the primary direction is equal to the neutrino
direction by definition, this correctly removes the effect of the
changing cross section of a flat detector. (Another viewpoint is that
the detector in a 1D calculation is simply a point, so the
flatness issue never arises).  In the 3D calculation,
$\theta_p \ne \theta_z$ in general and the flat detector problem has
to be addressed by introducing weights.  There are two choices which
have both been tried and give consistent results: (1) Generate
particles in the atmosphere in the ``natural'' way, i.e.\ proportional
to $\cos\theta_p$, and when each neutrino strikes the detector, it
receives a weight of $1/\cos\theta_z$ to compensate for the reduced
cross section; (2) generate particles isotropically (uniformly in
$\cos\theta_p$) and weight each neutrino which strikes the detector by
$\cos\theta_p/\cos\theta_z$.  The second method has the advantage that
in the limit of high neutrino energy where the 3D calculation generates
essentially collinear showers, the technique tends to the 1D technique
and the weight tends to unity.  The second method is used in the
calculations presented in this paper.

Both methods have the drawback that as the neutrino direction
approaches the horizon the weights diverge.  This reflects that fact
that the probability of counting such an event in the Monte-Carlo
tends to zero.  This can lead to large fluctuations in the flux
computed near the horizon, a feature which is unacceptable for use in
experimental analysis of underground detector data to extract neutrino
oscillation parameters.  We have studied various techniques for
avoiding large fluctuations~\cite{TsukubaTech} and have adopted a
`binlet' weighting scheme for this calculation.  As each neutrino
arrives at the flat detector, a bin in $\cos \theta_z$ is assigned to
it.  All neutrinos which enter this bin are assigned the weight given
by the value of $\cos\theta_z$ at the center of the bin rather than at
the individual value for each neutrino.  It is important to use narrow
bins for this to avoid bias, hence the name `binlet'.  For the present
analysis, 80 `binlets' are used in the range $-1<\cos\theta_z<1$, so
the maximum variation in weights is slightly less than 1:80.  Further
details of this and other viable techniques are discussed in
Ref.~\cite{TsukubaTech}.

To perform the `binlet'-weighting, a special histogramming package was
written.  This also allows the squares of the weights to be
accumulated for computation of the Monte-Carlo statistical error on
each point.  These errors are shown on the figures in this paper.  

\begin{figure}[htb]
\begin{center}
\epsfig{file=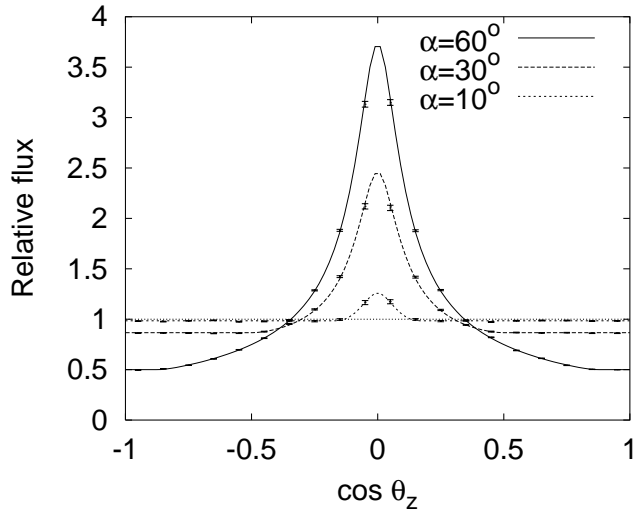,height=7cm}
\end{center}
 \caption{Comparison between a test Monte-Carlo calculation (points)
 which introduces a single bend through an angle $\alpha$ at an
 altitude of $h$ and an analytic calculation from~\protect\cite{Lipari1}
 (lines) for different values of $\alpha = 10^\circ$, $30^\circ$ and
 $60^\circ$.  These curves show the basic features of a 3D calculation
 (see text).}
\label{fig:analytic}
\end{figure}

To verify the functioning of the 3D calculation chain, we replaced the
hadron production generator by a simplistic model which generated a
neutrino with a fixed angle $\alpha$ to the primary, instantly at the
injection sphere where the primaries were injected.  This is shown in
figure~\ref{fig:analytic} for various different injection heights,
compared with analytic formulae from Ref.~\cite{Lipari1}.  The figure
illustrates the basic expectations of the 3D geometric effects - a
sharp enhancement near the horizon and a small reduction in flux near
vertical.  The effects increase strongly with increasing $\alpha$ and
are therefore expected to be largest for low neutrino energy where
interaction and decay of particles causes the largest angular change
between primary and neutrino directions.

\section{Results}

The three dimensional simulation proceeds by running at each energy until
about 1.2 million neutrinos hit the detector (or
until one million primaries have been generated).  The latter limit
only occurs at low energy.  All neutrinos passing within a circle of
radius 500 km centered on the detector location are accepted.  The
calculation has been done so far at the site of the Super Kamiokande
detector (Kamioka, Japan; Lat: 36.42~N, Lon: 137.310~E, Alt: 372~m)
and at two adjacent northern latitude sites, Soudan-2/MINOS (Soudan,
Minnesota, USA; Lat: 47.822~N, Lon: 267.752~E, Alt: Sea level) and SNO
(Sudbury, Ontario, Canada; Lat: 46.475~N, Lon: 278.632~E, Alt:
-1.7~km).

Several modes of operation of the code are available to allow the
change between a full 3D calculation and a 1D calculation to be made
in steps in the same program.
\begin{itemize}
\item {\bf 3D} The full 3-dimensional treatment.
\item {\bf NM} The 3D treatment, but with bending of particles within 
      the atmosphere turned off.
\item {\bf pseudo-1D} All transverse momenta are set to zero and 
      there is no bending within the atmosphere.   Particles are still 
      generated in all directions over the entire globe and the 
      details of the flat detector in three dimensions are still
      included.  In principle, running in this mode should produce 
      results in agreement with a 1D calculation, allowing us to test 
      the integrity of the 3D cascade code in the absence of 3D effects.
\item {\bf 1D} It is also possible to operate the 3D program in a 
      classical 1D mode, in which there is no $p_T$, no magnetic field 
      and only trajectories pointing directly at the detector center 
      are generated.  This differs from the pseudo-1D calculation 
      in that the weighting schemes are not needed.  
\item {\bf original-1D} The original code from Ref.~\cite{AGLS} was used
      with the 2.1 version of target.
\end{itemize}
Comparisons among these five calculations are shown in the
following part of the paper.  To summarize, original-1D, 1D and
pseudo-1D agree with each other in all distributions except for
statistical fluctuations in the pseudo-1D.  This agreement is
important as it checks the correct realization of the weighting
procedures in the pseudo-1D calculation, which are identical to those
in the 3D.  Generally NM agrees with either the 3D or the 1D runs as
described below and gives insight into the origin of the differences
between 1- and 3-dimensional calculations.

\begin{figure}[htb]
\begin{center}
\epsfig{file=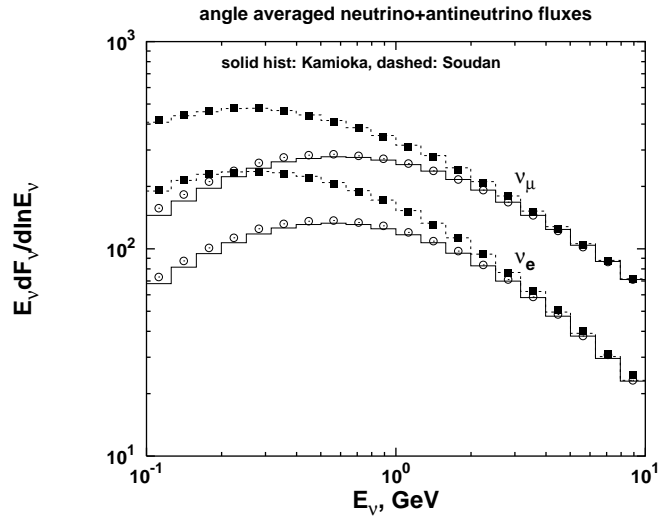,height=7cm}
\end{center}
 \caption{Comparison of 3D to 1D calculation of the angle averaged
 fluxes at Kamioka and Soudan.  The $(\nu_\mu+\overline{\nu}_\mu)$ 
 and $(\nu_e+\overline{\nu}_e)$ fluxes are plotted for the 3D 
 calculation (points) and the 1D calculation (lines).}
\label{fig:compe}
\end{figure}

The angle integrated fluxes of neutrinos and antineutrinos at
Kamioka and Soudan are compared with each other in
figure~\ref{fig:compe}.  The fluxes at SNO are similar to those at
Soudan.  The 3D and original-1D fluxes are plotted.  All three 1D
calculations are identical on this scale over the whole energy range
between 0.1 and 10 GeV.  The fluxes are before any oscillations, 
and the calculations were performed in the epoch of solar minimum.  
The large difference in fluxes between Kamioka and Soudan is due to the
difference in local cutoffs; Kamioka is near the geomagnetic equator
where the downward cutoffs are large (around 20~GeV, but direction
dependent) while Soudan is near the geomagnetic pole where the
downward cutoffs are only a few GeV and more low energy primaries get
through.  This plot shows that in changing from 1D to 3D calculation, 
there is a $\sim 3$\% increase in 3D sub-GeV angle-averaged fluxes.
For $E_\nu>1$~GeV, the angle-integrated fluxes from the 3D and 1D
calculations are identical within statistical differences of about 
a per cent.

\begin{figure}[phtb]
\begin{center}
\epsfig{file=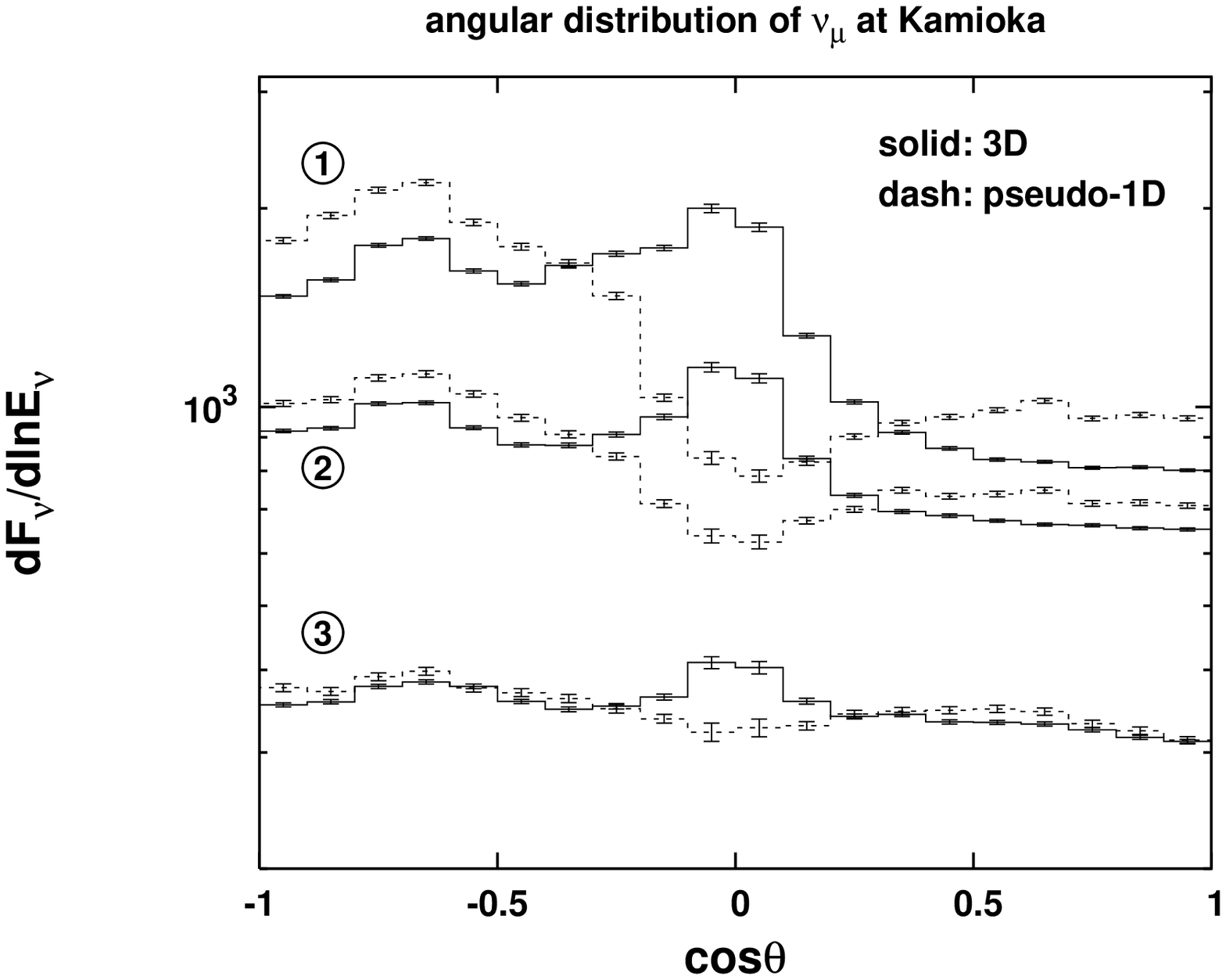,height=6cm}
\epsfig{file=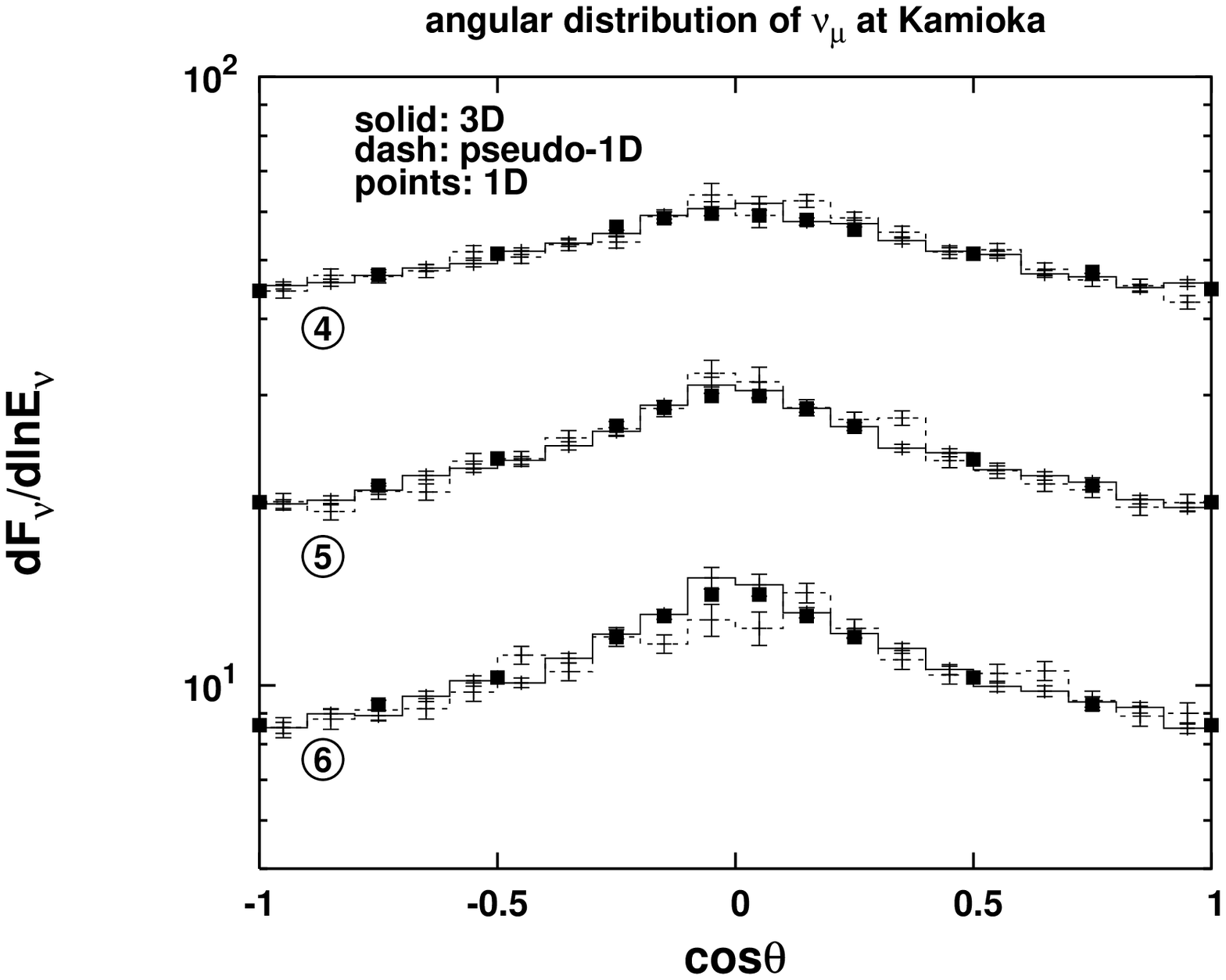,height=6cm}
\end{center}
 \caption{Zenith angle distributions of $\nu_\mu$ at
 Kamioka for six different energy ranges.  The full-line histograms
 are the 3D calculation and the dashed histograms are the pseudo-1D
 calculation.  The energy ranges in the left hand panel are:
 1) 100--158~MeV, 2) 250--400~MeV, and 3) 630~MeV--1~GeV. In the right hand
 panel we show the angular distribution for 4) 2.50--4.0~GeV, 5)
 4.0--6.3~GeV, and 6) 6.3--10~GeV. The right panel also contains the 
 angular distributions calculated with Bartol's original code (points).}
\label{fig:zenkam}
\end{figure}

Figure~\ref{fig:zenkam} compares the zenith angle distributions at
Kamioka (solar minimum) obtained with the 3D and pseudo-1D modes for
various energy ranges. It illustrates the second essential feature of
the change from 1D to 3D calculation, which was first noted by
Battistoni et al.\cite{Bat1}:
At low energy, the geometrical effect causes an
enhancement in 3D-fluxes near the horizon and a smaller depletion of
fluxes near the zenith.  The effect decreases as the neutrino energy
increases.  At higher energies, the two distributions become more
similar.  The broad peak in the high-energy flux is common
to 3D and 1D calculations.  It is caused by a combination of
two effects: decay of charged pions is enhanced
at large angles relative to hadronic interaction
($\sec\theta$ effect), and higher-energy muons at larger zenith angle can still decay
before reaching the ground.

At neutrino energies exceeding 5 GeV the two zenith angle
distributions become identical.  There are, however, still some
differences in the azimuthal distributions, which will be
discussed below. 

\subsection{Ratios of fluxes} 

\begin{figure}[phtb]
\begin{center}
\epsfig{file=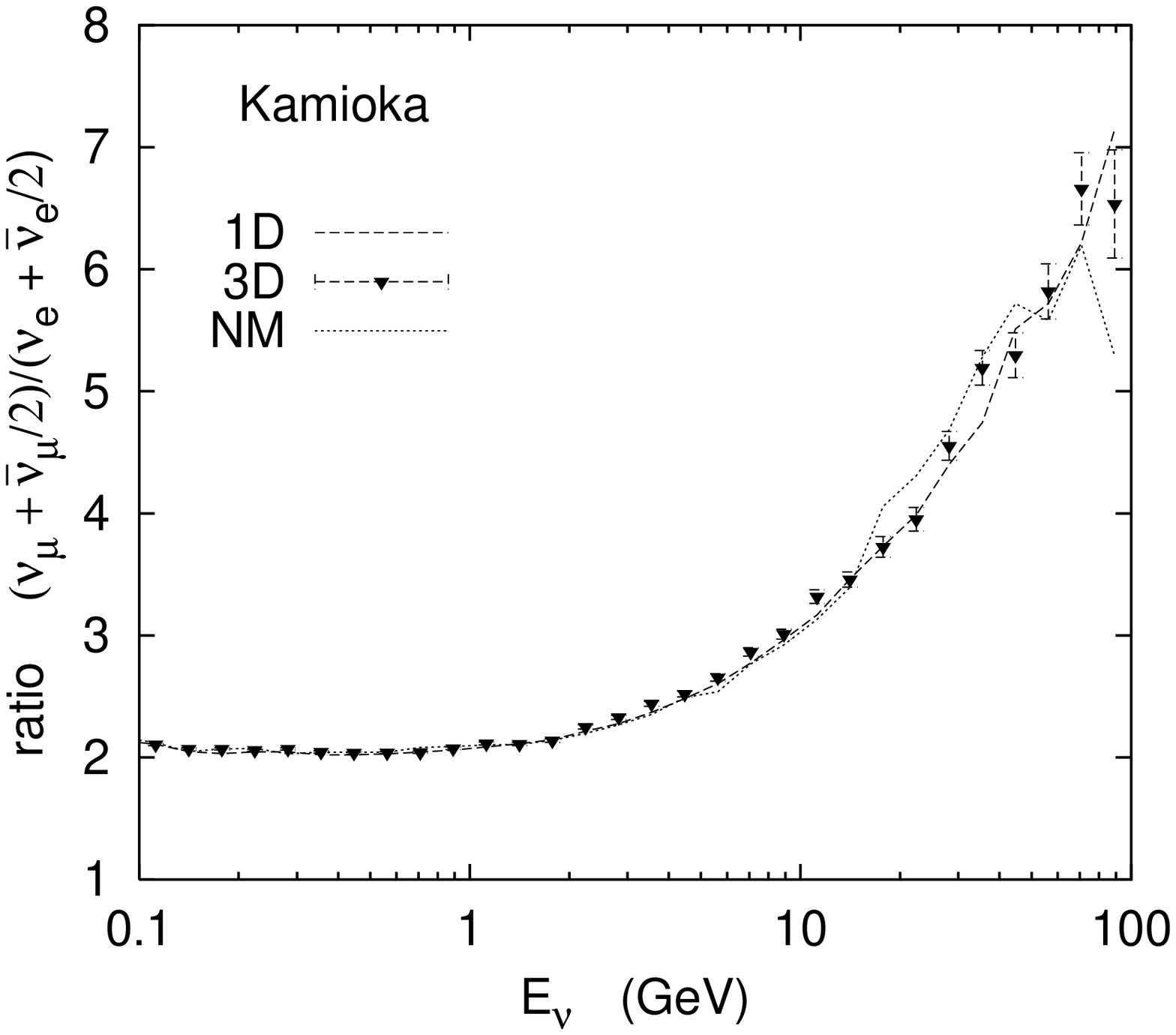,width=0.40\textwidth}
\epsfig{file=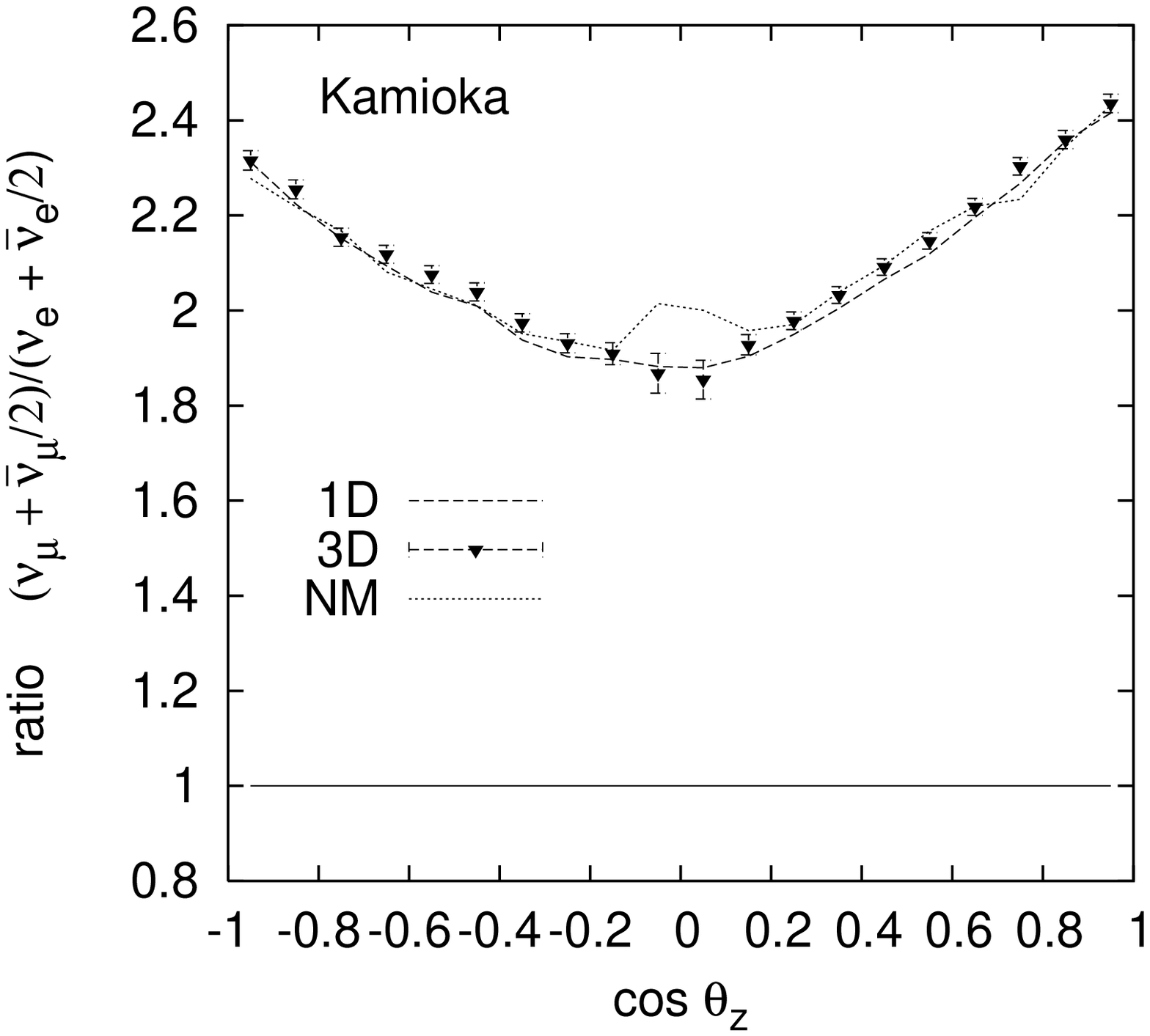,width=0.40\textwidth}
\end{center}
 \caption{Ratio of muon-like to electron-like neutrinos as a 
          function of (a) $E_\nu$ and (b) zenith angle 
          (for $E_\nu>315$~MeV), comparing 
          1D, NM and 3D distributions.  The error bars (not shown) 
          on the NM points are about a factor 1.8 bigger than 
          those shown on the 3D points.}
\label{fig:reme}
\end{figure}

\begin{figure}[ptb]
\begin{center}
\epsfig{file=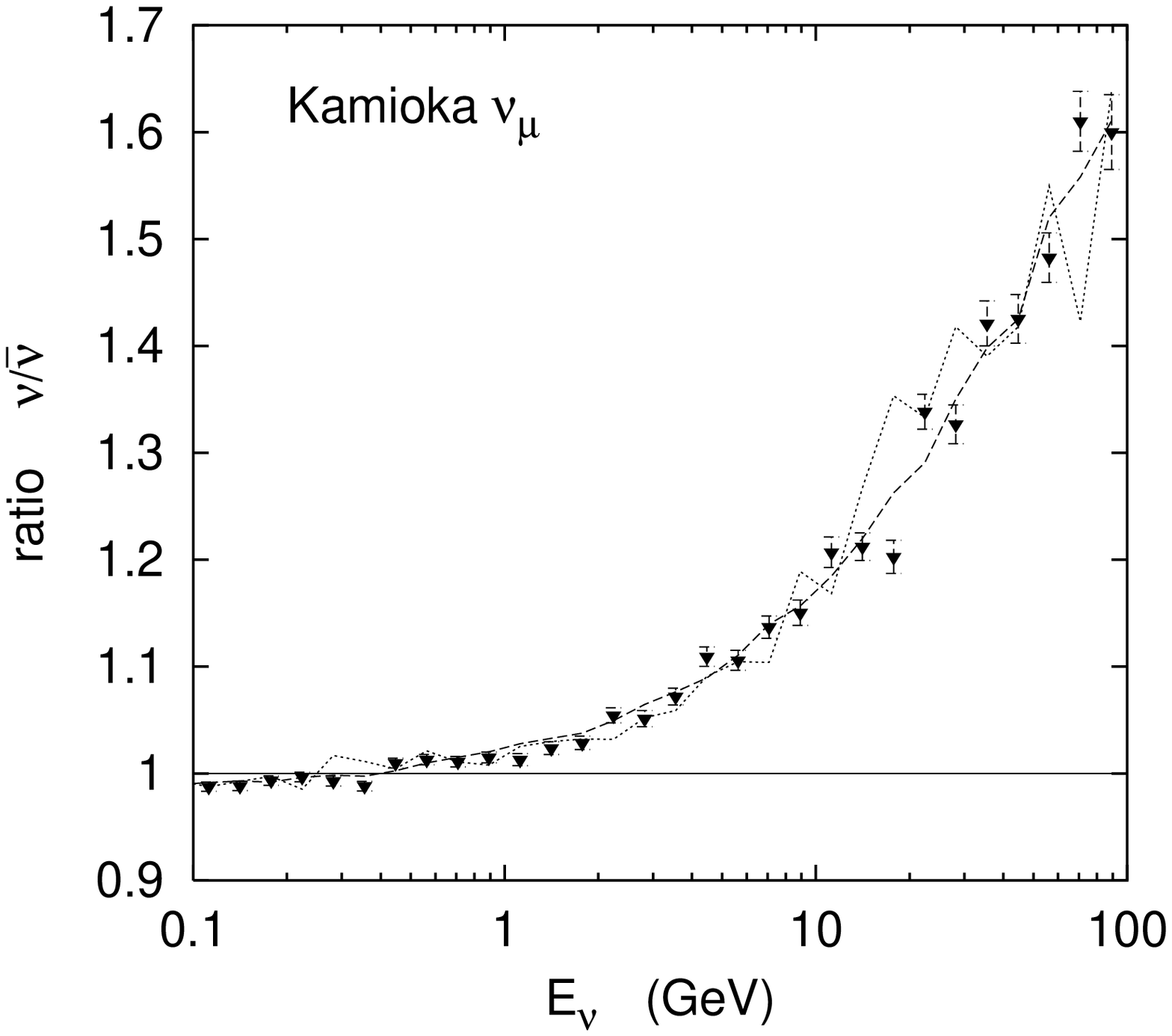,width=0.40\textwidth}
\epsfig{file=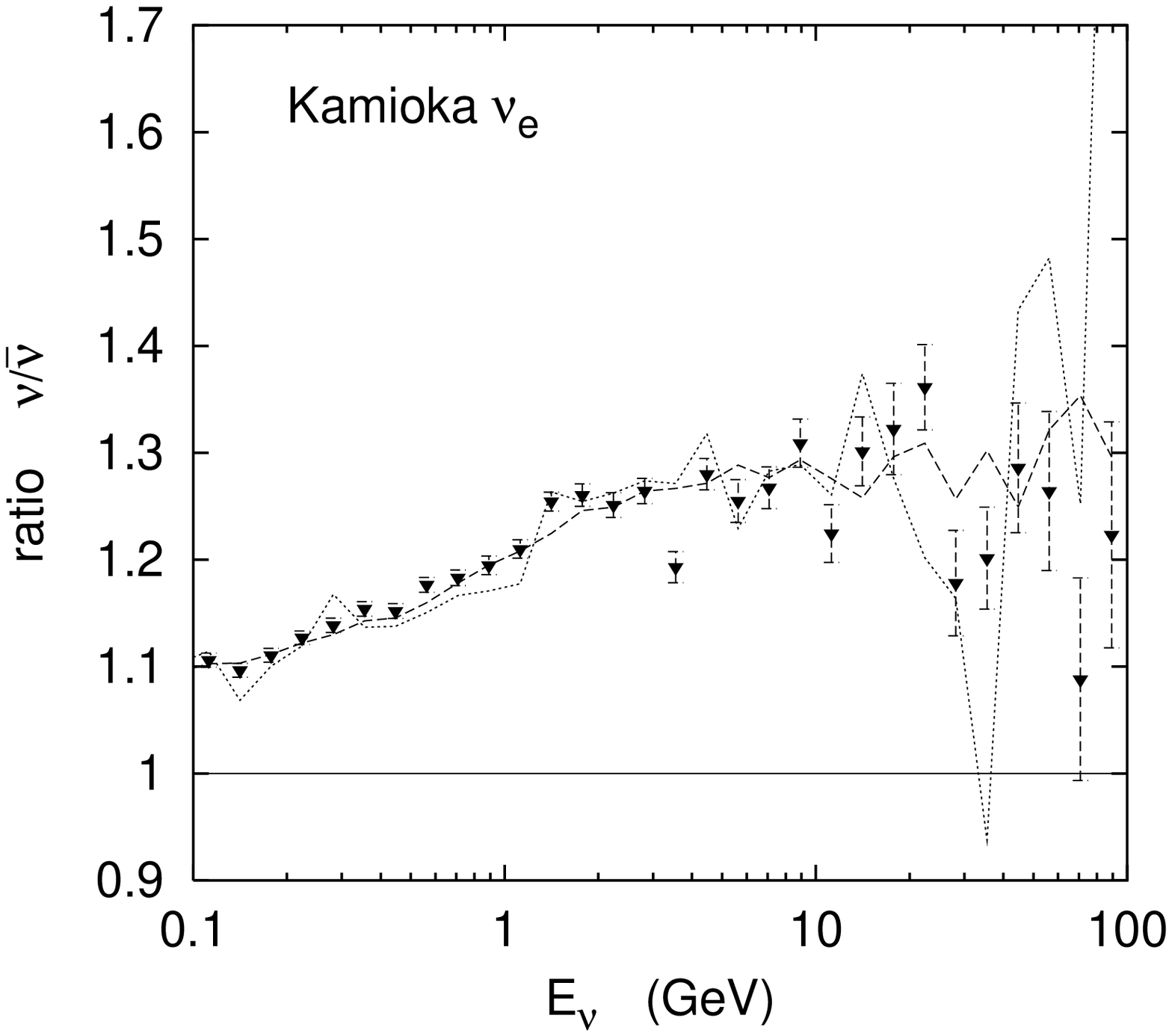,width=0.40\textwidth}
\epsfig{file=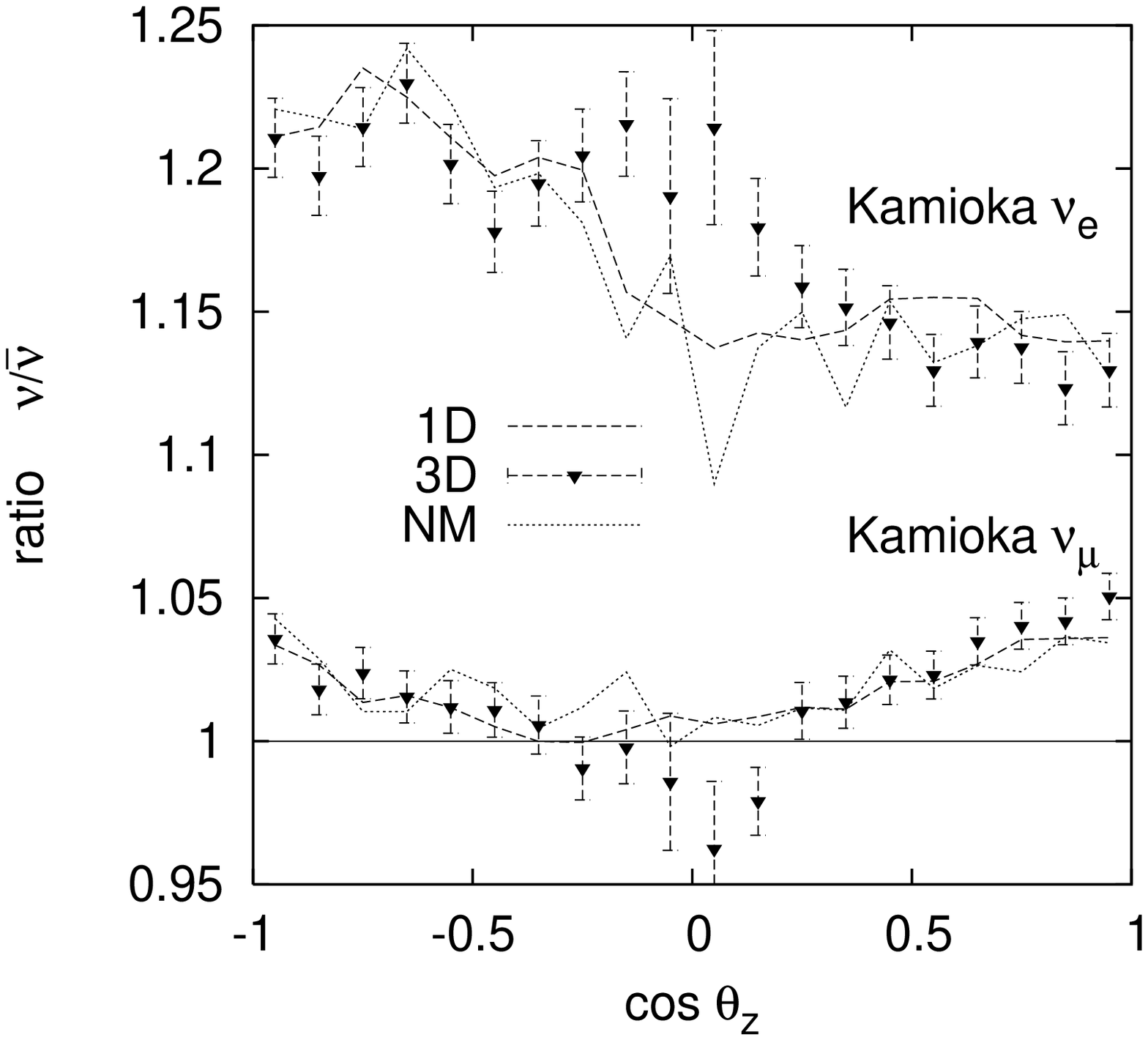,    width=0.40\textwidth}
\end{center}
 \caption{Neutrino to antineutrino ratios as a function of energy 
for (a) muon type neutrinos and (b) electron type neutrinos.  
(c) Zenith angle dependence of the neutrino to antineutrino ratios.}
\label{fig:rnb}
\end{figure}

\begin{figure}[htb]
\begin{center}
\epsfig{file=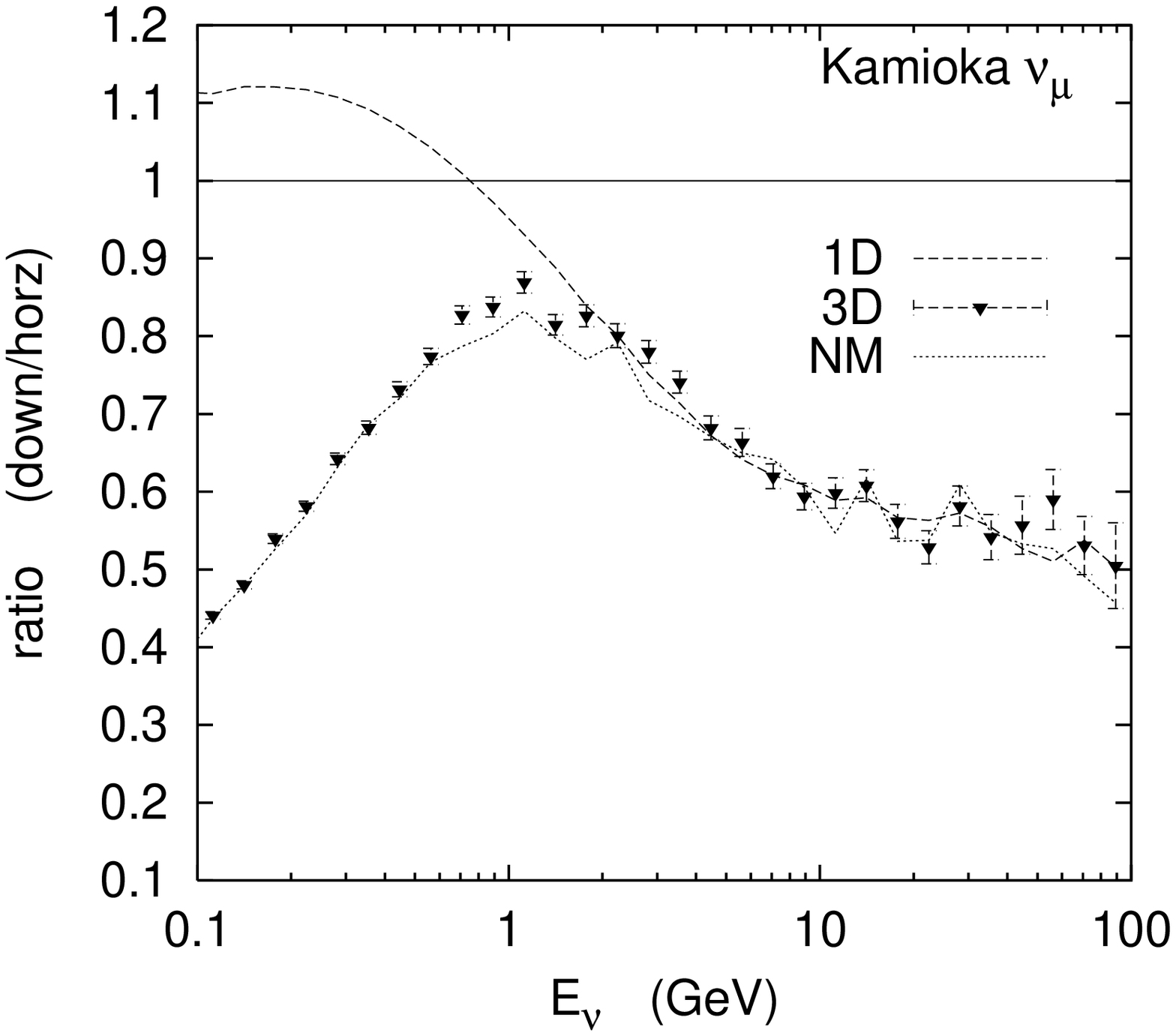,width=0.45\textwidth}
\epsfig{file=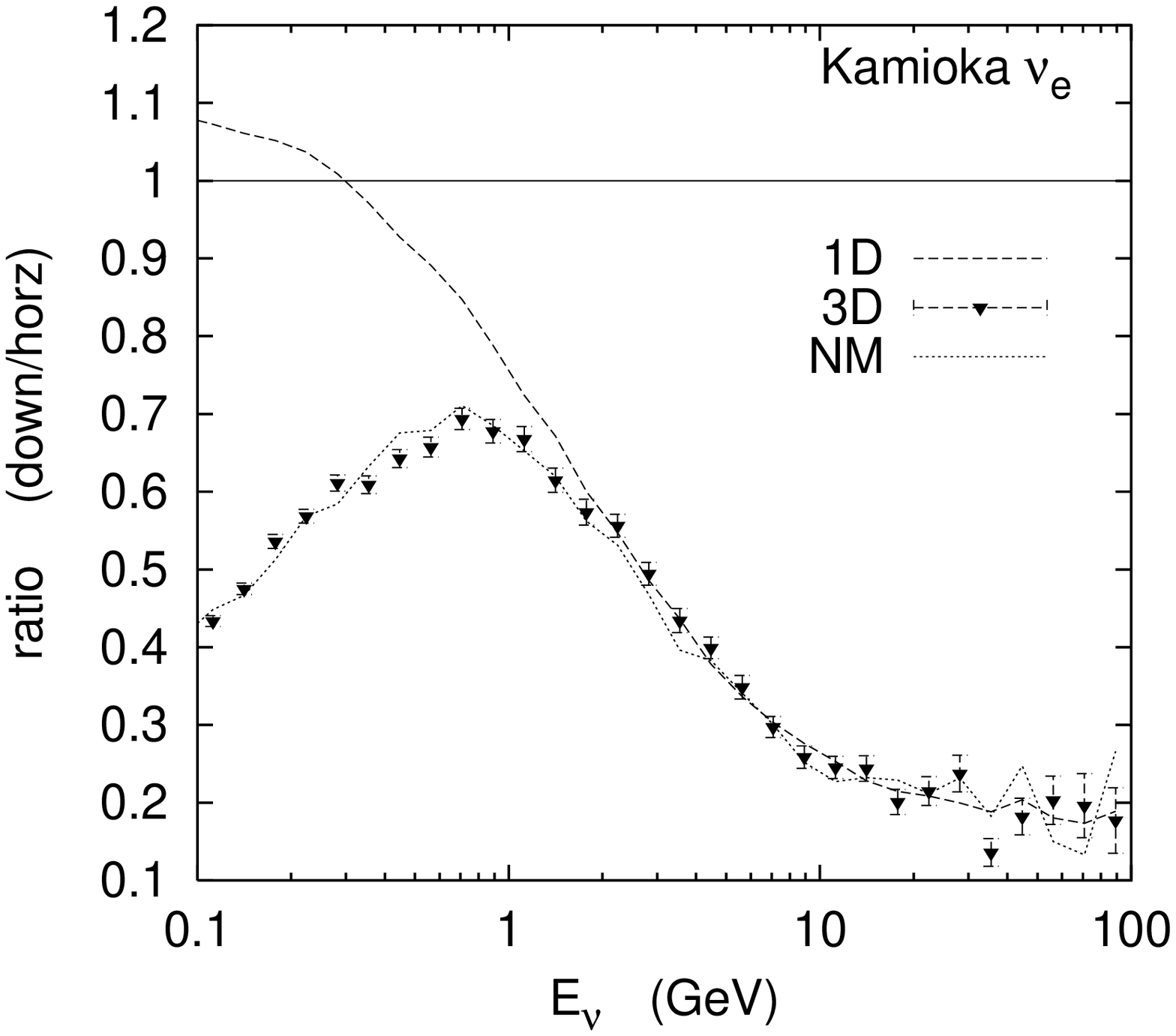,width=0.45\textwidth}
\end{center}
 \caption{Ratio of downward to horizontal neutrino fluxes (a) 
muon neutrino type and (b) electron neutrino type.}
\label{fig:dh}
\end{figure}

Figure~\ref{fig:reme} shows the ratio $R$ of muon-type to
electron-type neutrino fluxes as a function of neutrino energy for the
1D, NM and 3D calculations.  The ratio is calculated using 
$\nu\,+\,{1\over 2}\bar{\nu}$ as an approximate way to account for the difference in
interaction cross-section between neutrinos and antineutrinos.  The main features of
the ratio may be understood simply by counting the three
neutrinos (two muon-type and one electron-type) associated with each
muon in the atmosphere and noting that the kinematics works so that each
neutrino has about the same energy.  This gives $R = 2$.  At higher energies
$R$ increases as muons begin to reach the ground before decaying. 
There are no changes in the flavor ratio
between 1D and 3D calculations.  The fluctuation seen in the
zenith angle distribution near the horizon for the NM case is within
statistics.

\begin{figure}[htb]
\begin{center}
\epsfig{file=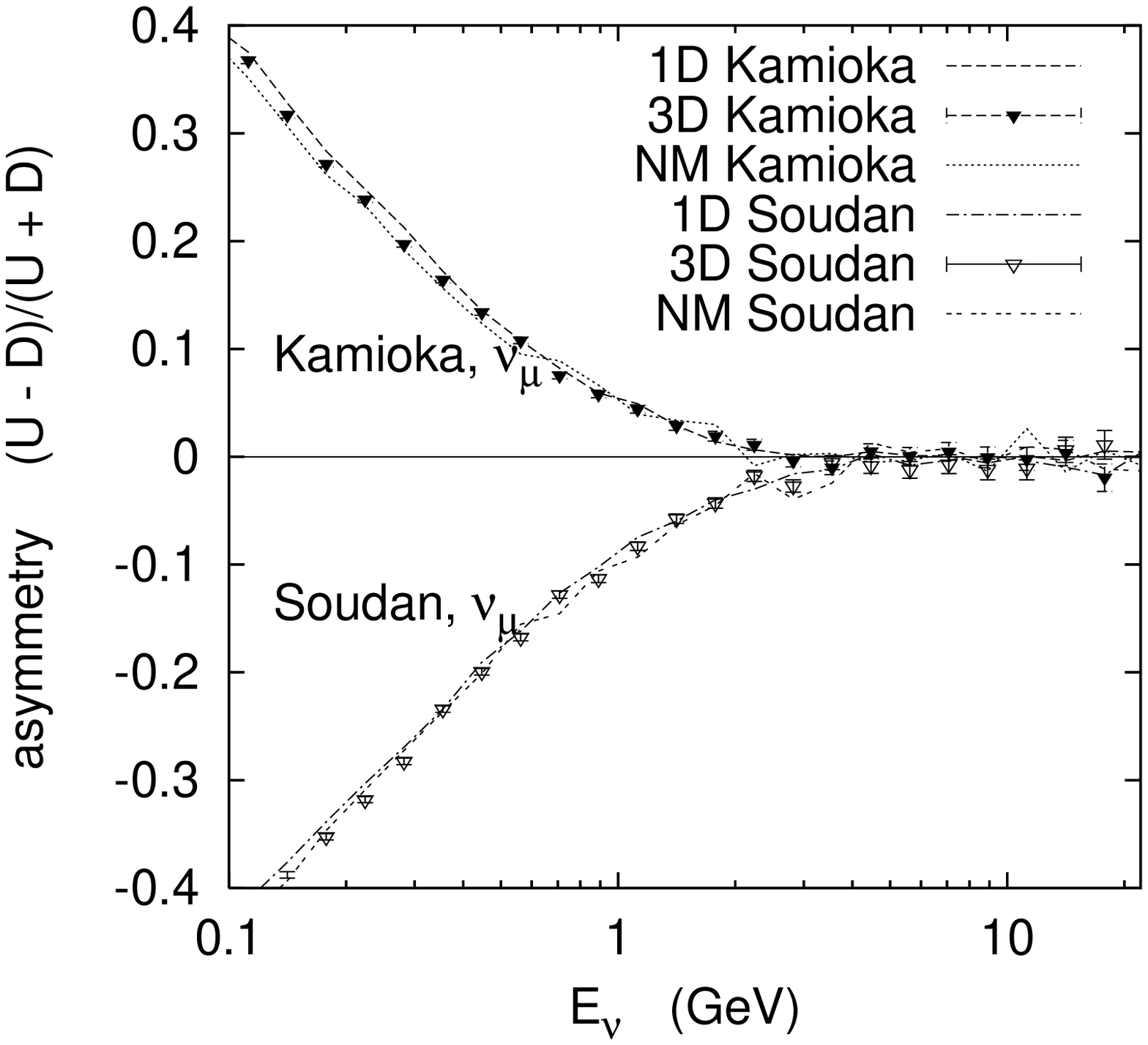,width=0.49\textwidth}
\epsfig{file=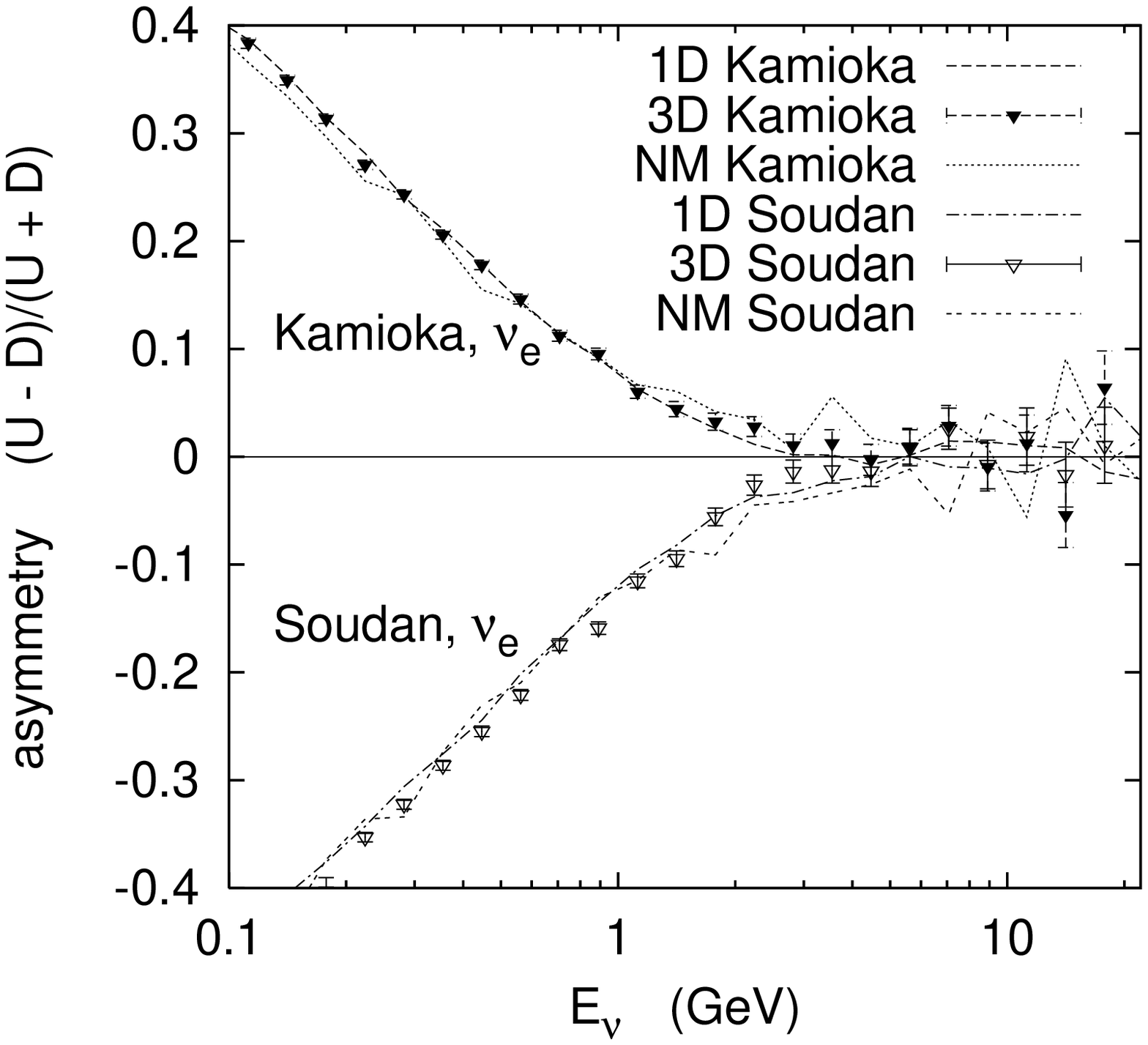,width=0.49\textwidth}
\end{center}
 \caption{Ratio of upward to downward neutrino fluxes as a 
function of neutrino energy plotted as an asymmetry for both
the Kamioka and Soudan sites. 
(a) muon neutrino type and (b) electron neutrino type.
}
\label{fig:uda}
\end{figure}

Figure~\ref{fig:rnb} shows the ratio of neutrinos to antineutrinos
as a function of energy and zenith angle.  At low energy, the
simple neutrino counting argument from above gives an expected ratio
of 1.  At higher energies where muons hit the ground, the ratios
increase a little to reflect the $\mu^+/\mu^-$ ratio of about 1.25
from hadron production.  Kaons play an increasing role in the
production at higher energies.  There are no changes in these ratios
between the 1D and 3D methods except for a possible slight excess in
the 3D above statistics near the horizon in the electron neutrino
ratio.  This is inconsistent with the much larger effect shown by
Wentz~et.~al~\cite{Wentz}.  Figures~\ref{fig:dh} and~\ref{fig:uda} show the
down to horizontal ($R_\mathrm{DH}$) and up to down ($R_\mathrm{UD}$)
ratios respectively (each of down, horizontal and up fluxes
are taken in an interval of 0.4 in $\cos(\theta_z)$).  $R_\mathrm{DH}$
deviates significantly between 1D and 3D at low energy due to the
geometric effect, as seen earlier on figure~\ref{fig:zenkam}. 
A conservative limit above which 3D effects can safely
ignored in calculating this ratio is $5$~GeV.  
The NM agrees with the 3D.  The up-down ratios are
consistent between 1D, NM and 3D at all energies.  The change in the
ratio with energy is governed by the difference in geomagnetic cutoff, which is
locally very high at the Kamioka site and low at northern sites
such as Soudan.

\subsection{Azimuthal distributions}

The azimuthal distributions of neutrinos are distorted by the curvature
of their charged ancestors in the geomagnetic field. Effects are most
important for primary cosmic rays and for the long-lived protons
and muons in the atmosphere.  As a consequence, 
the East-West asymmetry is different in
the results of a 3D calculation as compared to the 1D results.

\begin{figure}[phtb]
\begin{center}
\epsfig{file=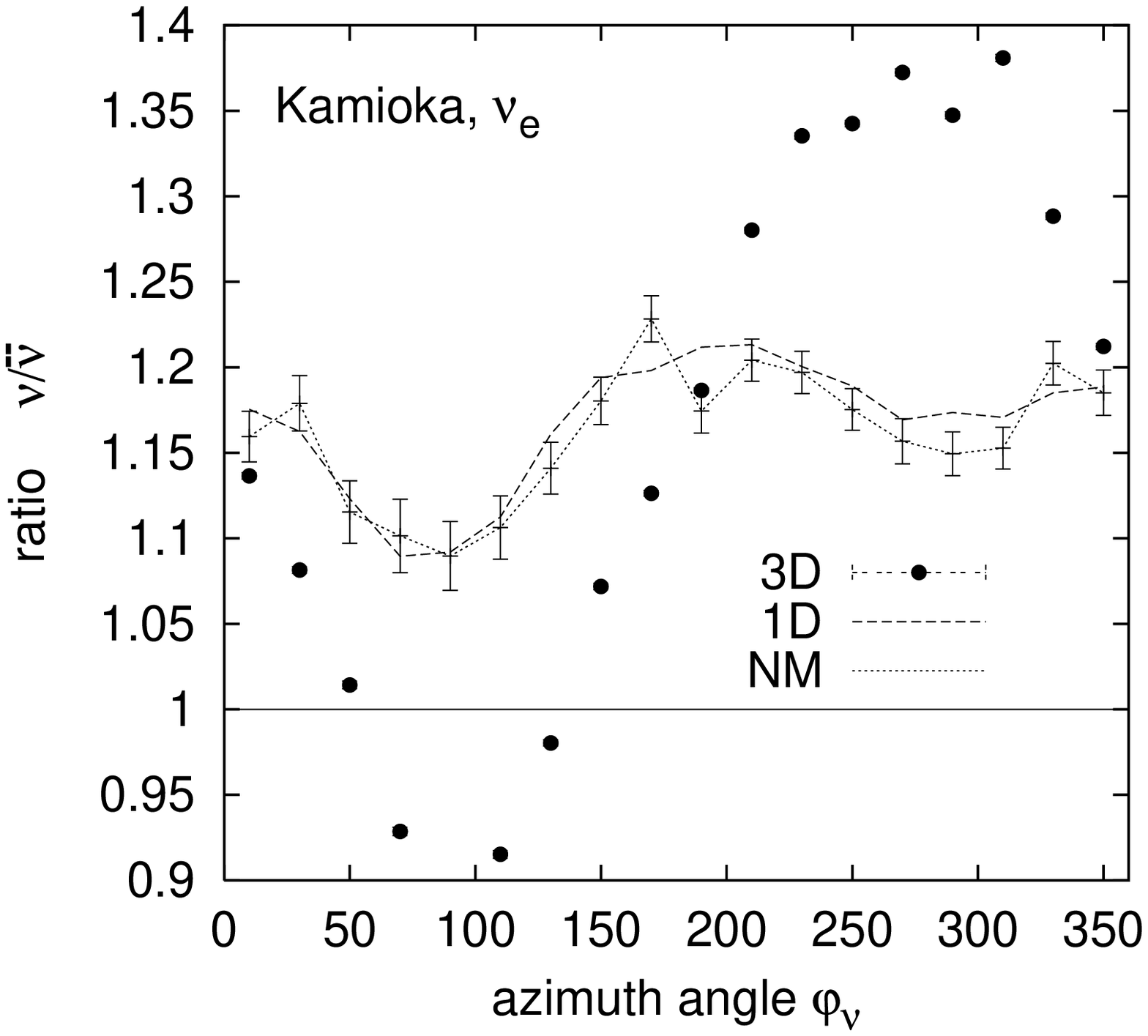,width=0.49\textwidth}
\epsfig{file=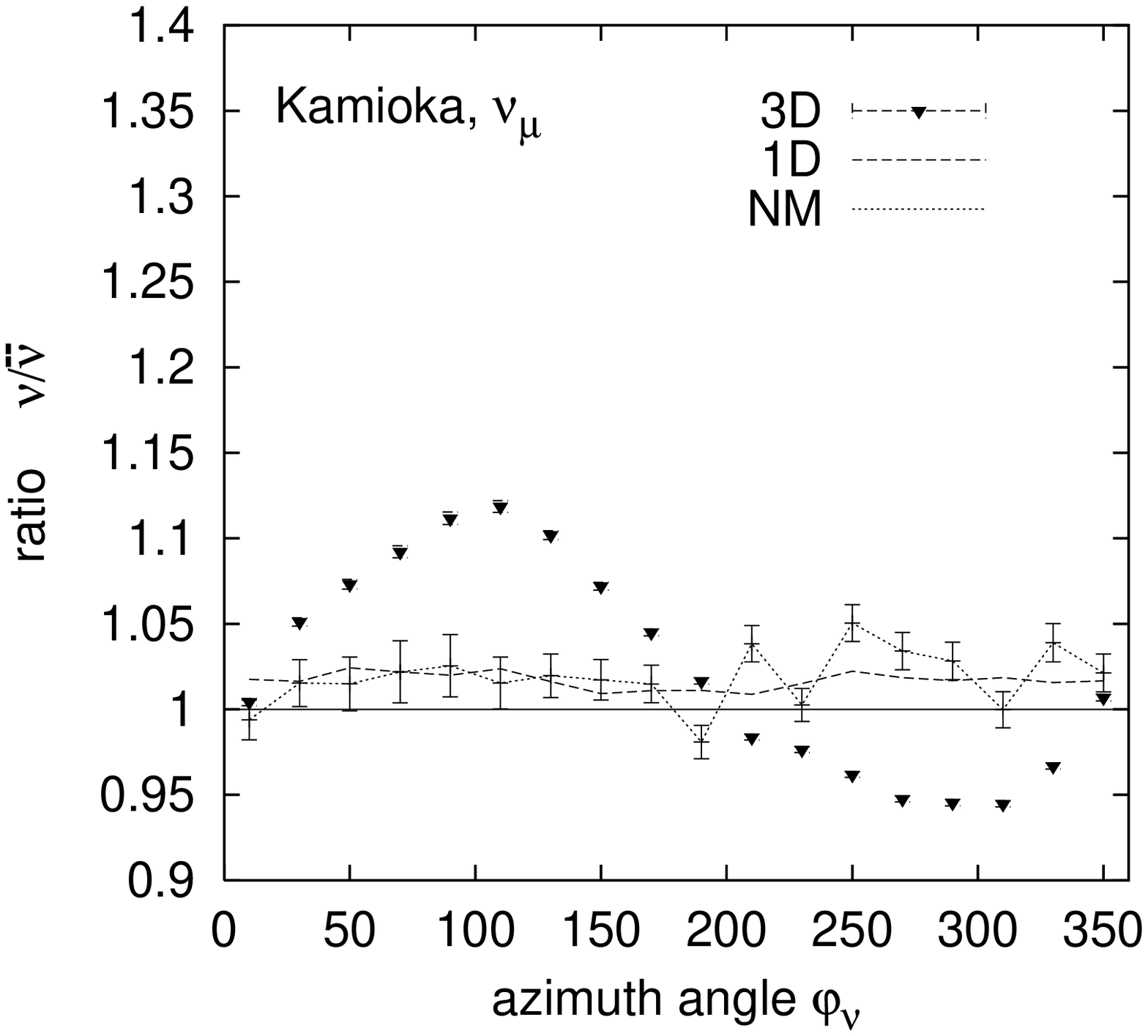,width=0.49\textwidth}
\end{center}
\caption{
 Neutrino to anti-neutrino flux ratios as a function of azimuthal angle 
 at Kamioka, for electron-like (left) and muon-like (right) neutrinos.
 Points: 3D calculation; dashed line: 1D calculation; dotted line (with 
 error bars): NM calculation (3D with the magnetic field neglected for 
 particle transport in the atmosphere).
}
\label{fig:azi4kam}
\end{figure}

The ratio of neutrino to anti-neutrino fluxes is shown as a function of 
the azimuthal direction $\varphi$ in figure \ref{fig:azi4kam} for electron-like 
neutrinos (left) and muon-like neutrinos (right), for the 1D, 3D and NM 
calculations.  We have defined $\varphi$ such that $0^\circ$ is for neutrinos
arriving from the North, $90^\circ$ for neutrinos from the East and so on.
The plots were made for neutrino energies above $315$~MeV.
We divide neutrino by anti-neutrino fluxes
so that anisotropies resulting 
from the primary cutoffs should largely cancel.
This plot therefore displays
anisotropies primarily due to particle transport through the atmosphere.
The 1D and 3D calculations differ strongly in these 
residual azimuthal anisotropies.  The NM calculation agrees with 
the 1D result, indicating that these differences are due to the bending 
of particles (mostly the muons) in the atmosphere.

\begin{figure}[phtb]
\begin{center}
\epsfig{file=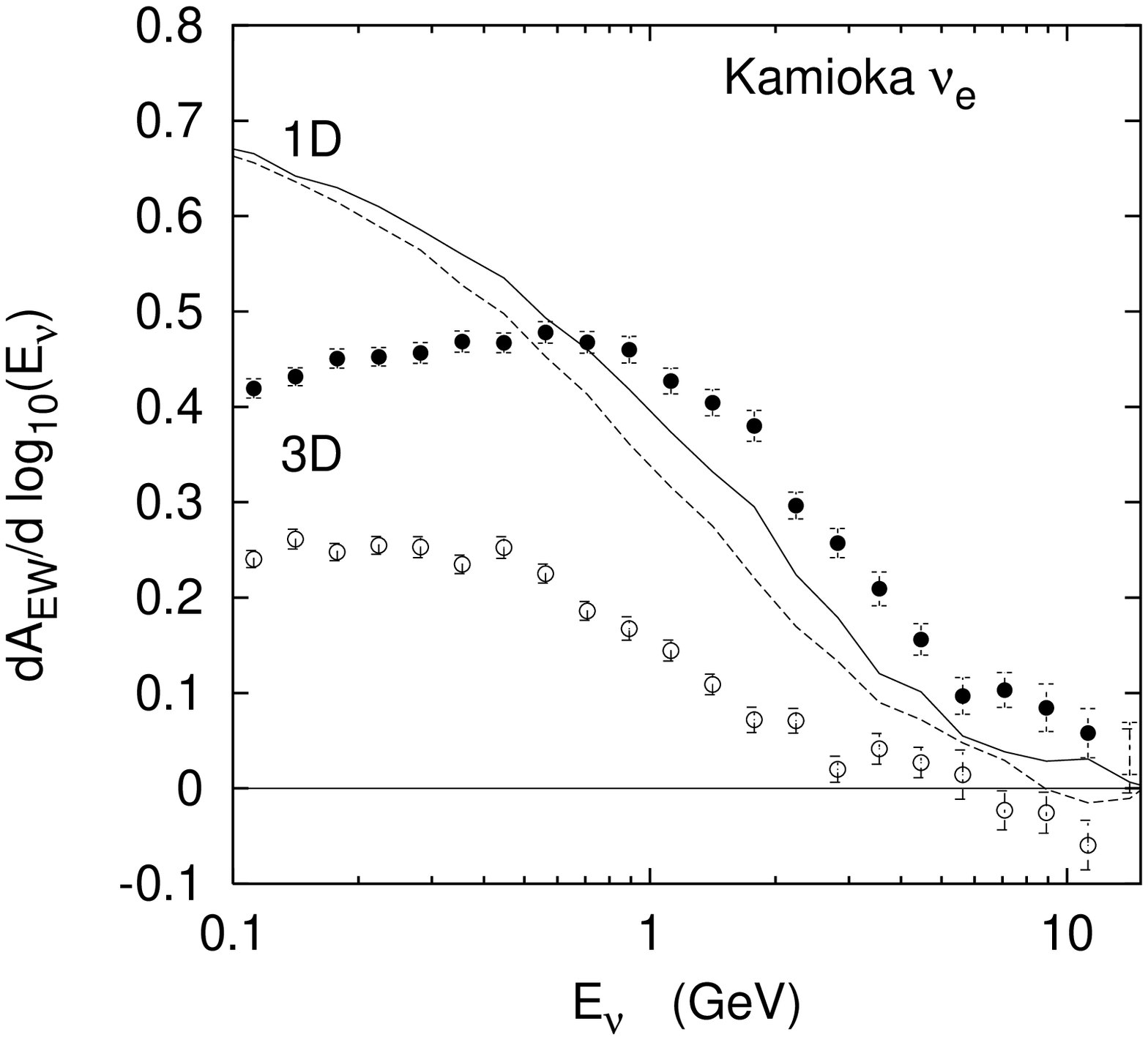,width=0.49\textwidth}
\epsfig{file=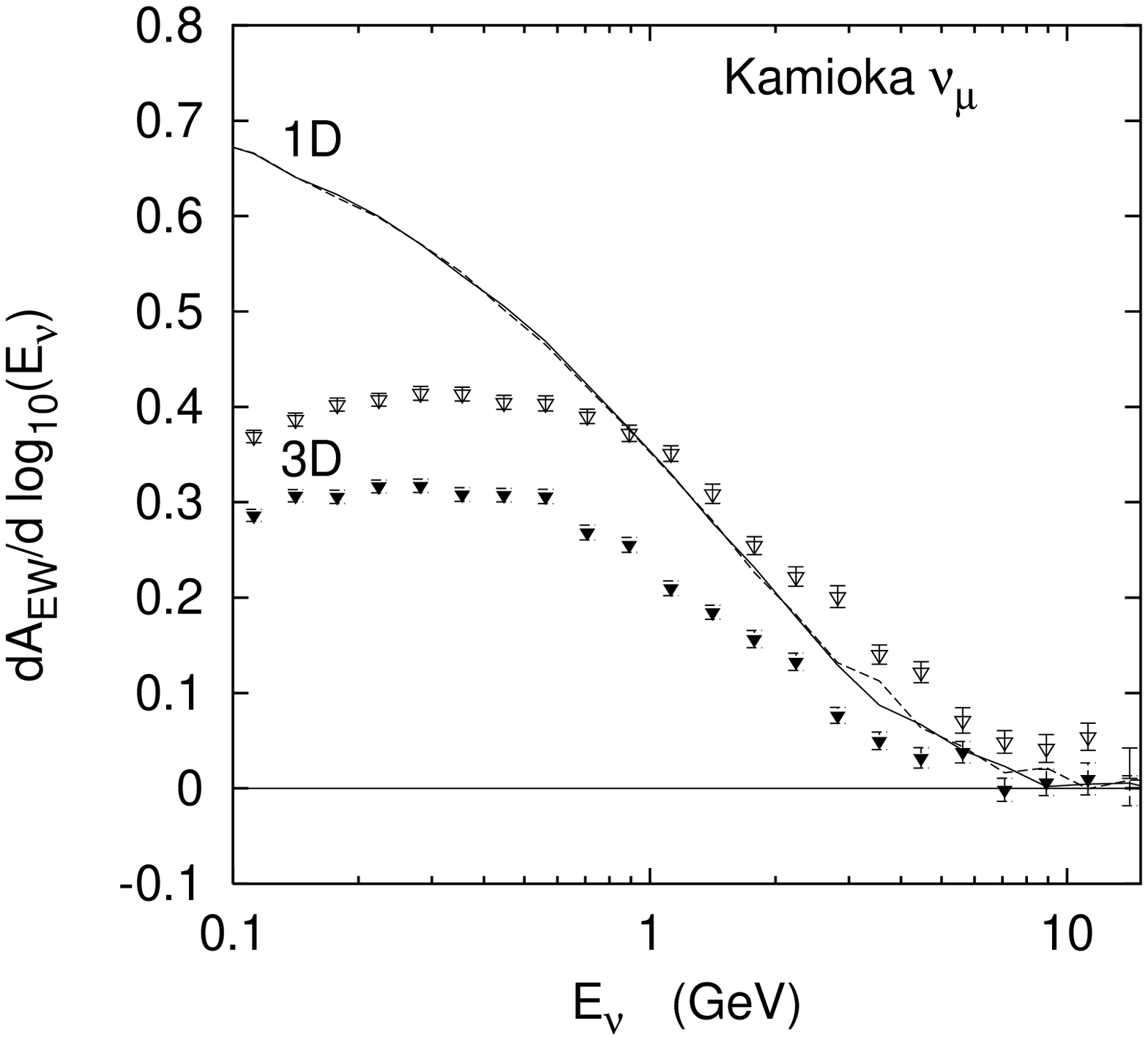,width=0.49\textwidth}
\end{center}
\caption{
 The size of the East-West asymmetry ($A_{EW}=(E-W)/(E+W)$) as a 
 function of neutrino energy for the four neutrino species: 
 $\nu_e$ (left, circles), $\nu_\mu$ (right, triangles); 
 neutrinos (filled, continuous line), anti-neutrinos (empty, dashed line).
 The lines show the results of the 1D calculation and the data points 
 the 3D calculation.
}
\label{fig:ew_e}
\end{figure}

The size of the East-West asymmetry $(E-W)/(E+W)$ for horizontal
neutrinos ($|\cos\theta_z|<0.5$) is shown in figure
\ref{fig:ew_e} as a function of neutrino energy for the 3D (points)
and 1D (lines) calculations.  East:
$40^\circ<\varphi<140^\circ$, West:
$220^\circ<\varphi<320^\circ$.  In this figure each of the 1D
lines agree well with each other, showing that the East-West effect is
the same for all neutrino types.  This is a consequence of the primary
cutoff, which is the same for all neutrino types and the only source
of azimuthal anisotropies in the 1D calculation.  The shape of the 1D
distributions is governed by the primary cutoff -- at low neutrino
energies the strength of the cutoffs is at a maximum, resulting in
the maximum East-West asymmetry; as the neutrino energy rises, so the
size of the asymmetry decreases as the primary cutoffs become
progressively less important.

The points in figure~\ref{fig:ew_e} show that, for low energy neutrinos,
the East-West asymmetry is much smaller for the 3D calculation; the NM
results (not shown) also agree with this.  This is because
the correlation between the neutrino and primary directions decreases
with decreasing neutrino energy, which is also related to the 3D
geometrical effect.  Thus, for low energy neutrinos the need for a
fully 3D calculation is clear; however, even at higher neutrino
energies the 1D and 3D results do not agree.
Both the bending of primary particles (between injection and
interaction) and muons is important to the East-West effect.  Whereas
the bending of protons produces the same result for all neutrino types
(it is an extension of the cutoffs) the bending of muons introduces
anti-particle/particle differences.  The effect due to muons is simply
a result of positive and negative muons bending in opposite directions
in the geomagnetic field, resulting in an enhanced East-West
asymmetry for the decay products of positive muons ($\nu_e$ and
$\bar{\nu_\mu}$) as shown in figure~\ref{fig:ew_e}.

The most important conclusion to draw from figure~\ref{fig:ew_e} is
that the East-West asymmetry is different in the 3D and 1D
calculations even for 10~GeV neutrinos, and that the size of this
effect is dependent on the neutrino type.

\subsection{Path length distributions}

\begin{figure}[htb]
\begin{center}
\epsfig{file=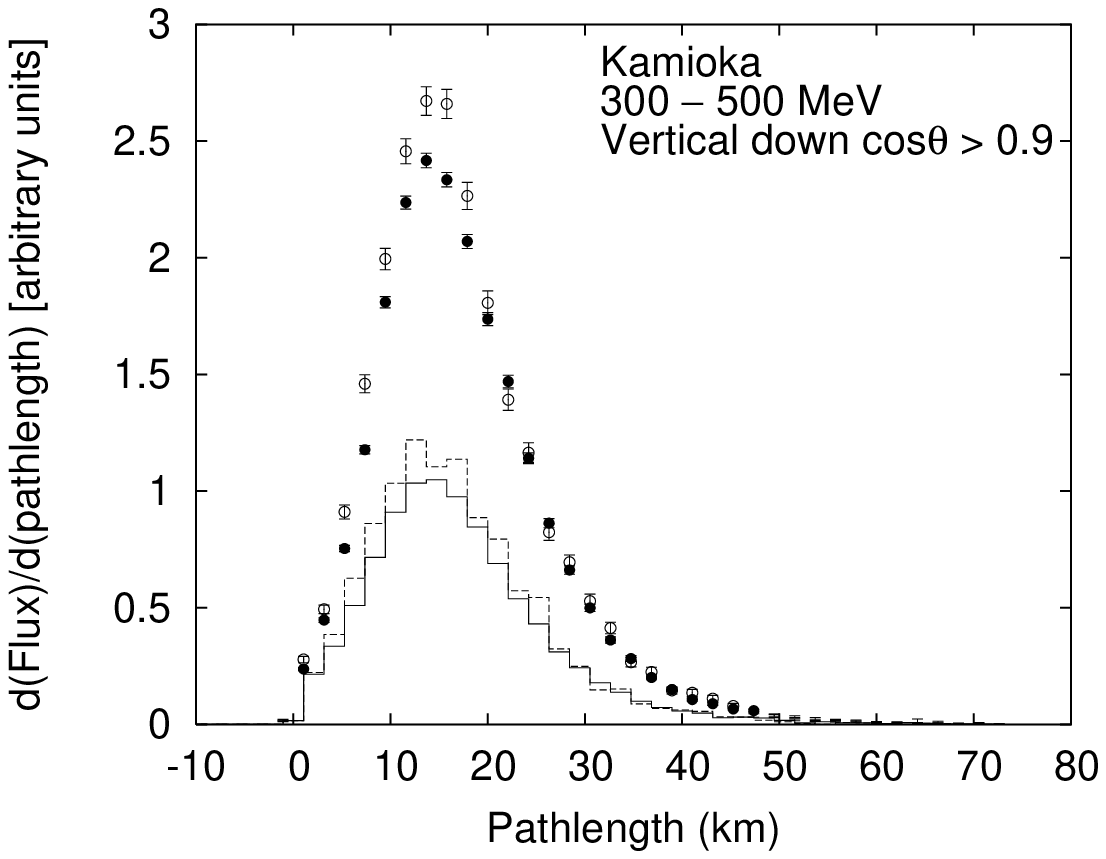,width=0.49\textwidth}
\epsfig{file=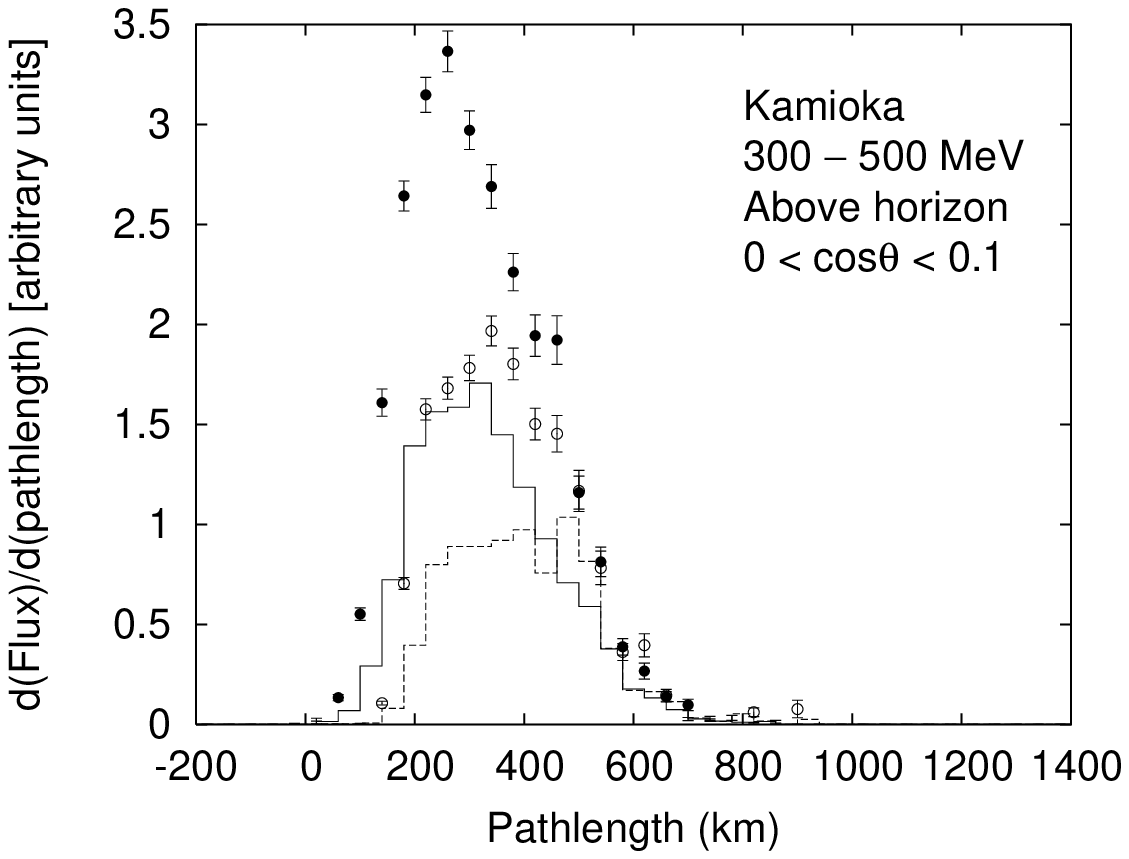,width=0.49\textwidth}
\end{center}
 \caption{Distributions of neutrino path length from production to
 detection for vertically downward neutrinos (left) and
 near-horizontal neutrinos (right).  Full (open) circles are muon
 neutrinos from the 3-D (1-D) calculation, and full (dashed)
 histograms are electron neutrinos from the 3-D (1-D) calculation.}
\label{fig:distdist}
\end{figure}

\begin{figure}[ptb]
\begin{center}
\epsfig{file=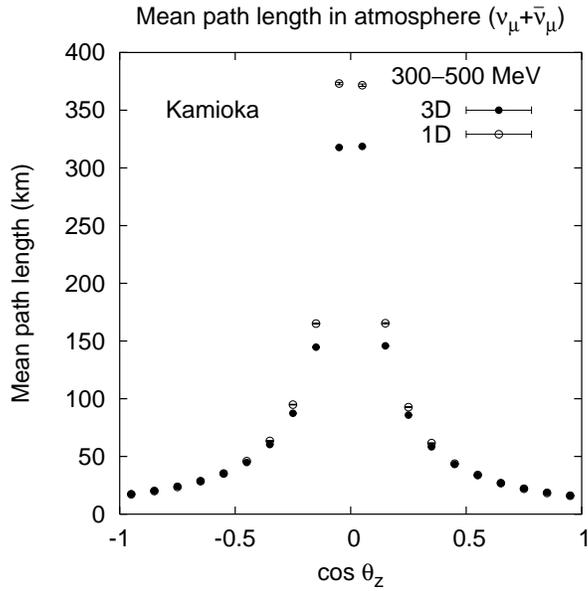,width=0.5\textwidth}
\end{center}
 \caption{Mean path length distance in the atmosphere of muon neutrinos
 and antineutrinos in the range 300--500MeV as a function of zenith
 angle, comparing 3-D and 1-D calculations.  For upward going
 neutrinos, the path length in the Earth must be added.}
\label{fig:meandistlow}
\end{figure}

The path length of the neutrinos from their production point to the
place where they are detected is an important consideration when
determining the neutrino oscillation parameters.  A difference is
expected between 1D and 3D calculations near the horizon.  In the 1D
calculation the only way of producing a horizontal neutrino is by the
interaction of a primary which grazes the atmosphere and (because of
the angle of incidence) interacts very high up.  In the 3D simulation,
there is the possibility that a horizontal neutrino may be produced
from a more vertical cascade with a neutrino emitted sideways -- such
neutrinos are generally produced closer to the detector.  This is
illustrated in figure~\ref{fig:distdist} which shows the distance
distribution for neutrinos produced vertically and near the horizon
with the 3D and 1D distributions superimposed on the same plot.  The
vertical distributions are nearly the same for 1D and 3D calculations
whereas the horizontal distributions have an additional component in
the 3D calculation at low path length (corresponding to horizontal
neutrinos being produced from more vertical cascades).  The mean
path length is shown as a function of zenith angle in
figure~\ref{fig:meandistlow}.  For upward going neutrinos, the
path length in the Earth ($2R_\oplus|\cos\theta_z|$) must be added.

\begin{figure}[ptb]
\begin{center}
\epsfig{file=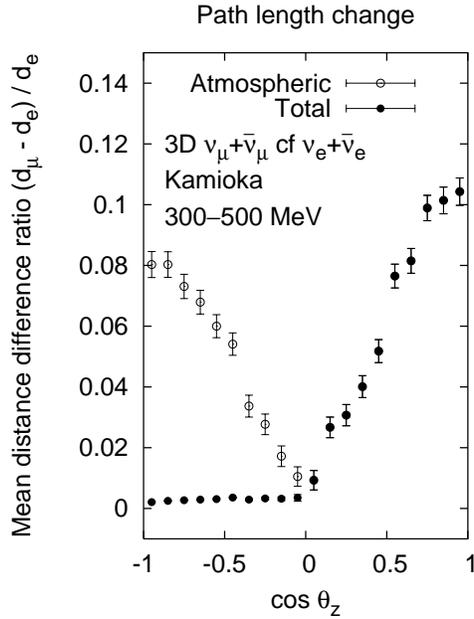,width=0.40\textwidth}
\end{center}
 \caption{Comparison of the path lengths of electron and muon type 
 neutrinos as a function of zenith angle.  Full circles compare the 
 total distances traveled (Earth + atmosphere) and open circles compare 
 the distance traveled in the atmosphere only.}
\label{fig:distme}
\end{figure}

The geometrical effect in the 3D calculation which causes the
enhancement in the flux near the horizon also causes the mean
production distance to be reduced.  The reduction is 15\% at the
horizon for neutrinos with energies in the range 300-500~MeV, and
quickly diminishes (2\% in the $0.4 < \cos\theta_z < 0.5$ bin).  At
higher energy also, the effect diminishes: the path length reduction at
the horizon is 4\% for neutrinos above 1~GeV.  Note that changing the
altitude of the detector (see following section) has
a similar sized effect on the path length distribution, it is important
that the true altitude of the detector is used and not just an
assumption that it is at sea level.

Figure~\ref{fig:distme} compares the path length distribution for
electron and muon type neutrinos.  Since the majority of electron
neutrinos are produced in muon decay while most muon neutrinos are
associated with muon {\it production}, muon neutrinos are produced on
average higher in the atmosphere - this effect is only important for
downward, nearly vertical neutrinos.

\subsection{Detector size}

An important technical question for a 3D calculation is, How large
can the ``detector'' be without distorting the results?
Since the geomagnetic cut-offs change rapidly at small
geomagnetic latitudes, such as in Japan, the adoption of a large
neutrino detector can, in principle, lead to incorrect predictions of
the atmospheric neutrino flux at such locations.  To study this
uncertainty, we have repeated the calculation moving the detector away
from its real location.

\begin{figure}[htb]
\begin{center}
\epsfig{file=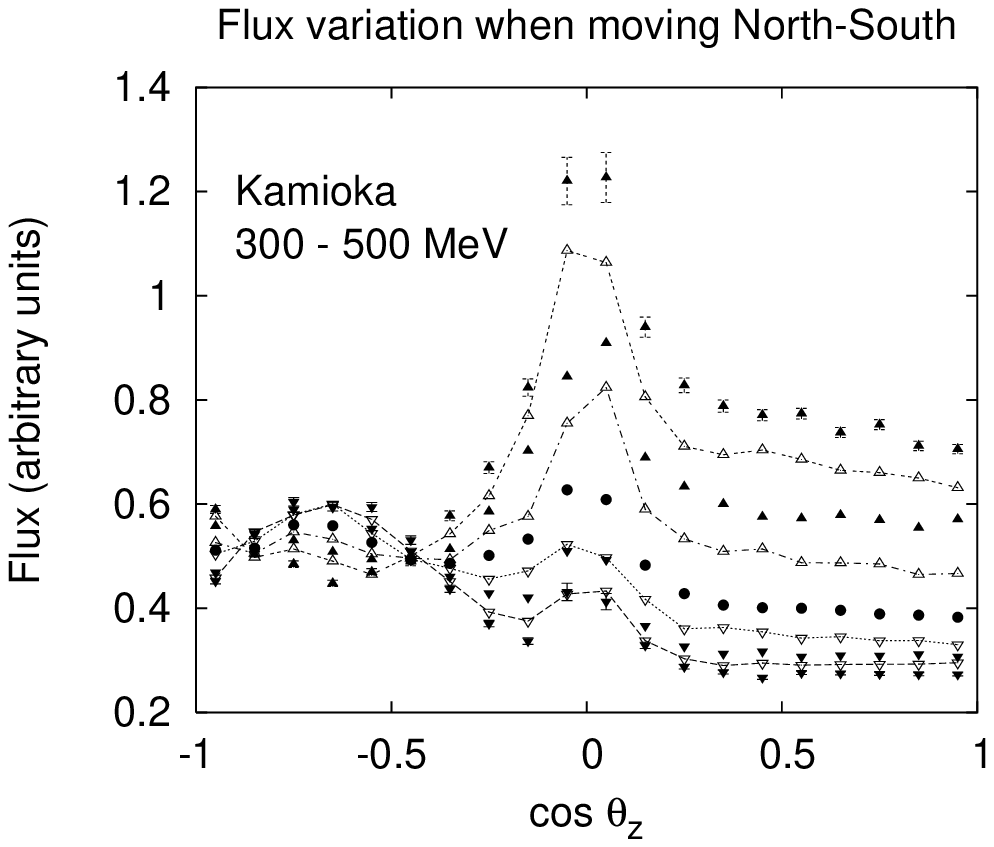,width=0.49\textwidth}
\epsfig{file=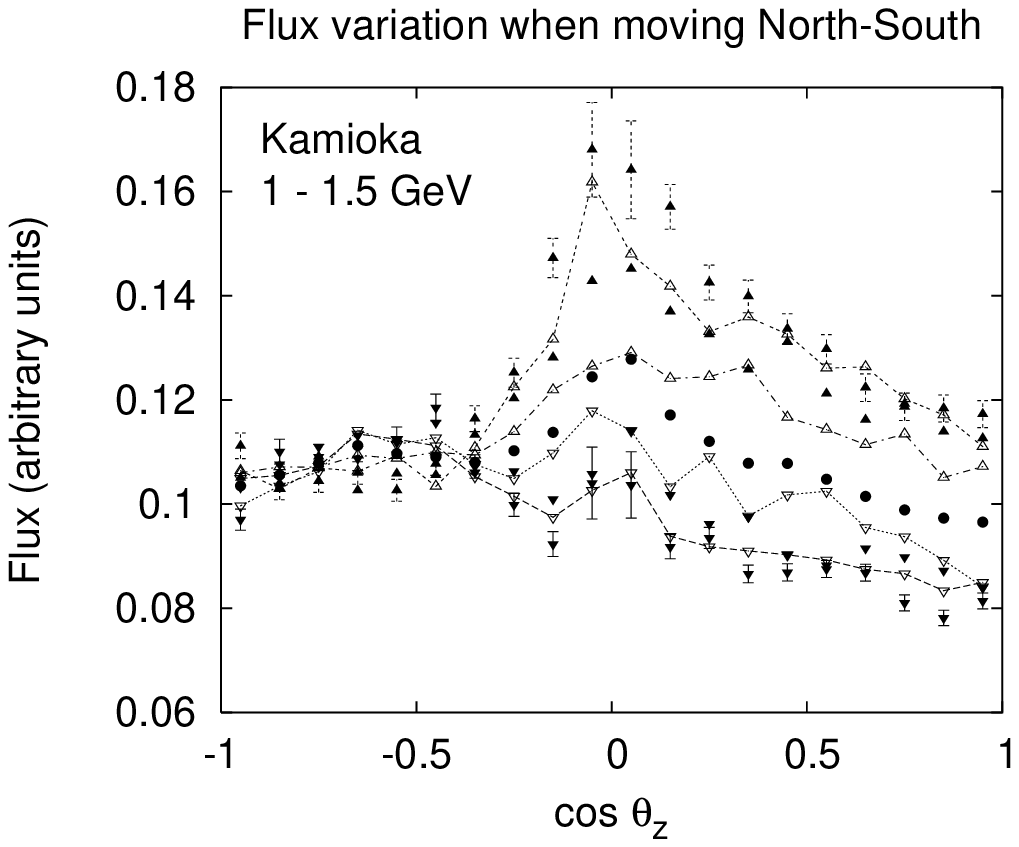,width=0.49\textwidth}
\end{center}
 \caption{Results of runs spaced 500 km apart in latitude.  The solid
circles are at the Kamioka site, upward triangles are for runs going
north from the detector site (alternating full and open symbols) and
downward triangles are for runs going south.  The lines are shown to
guide the eye by joining the points on every second curve (at
$\pm$500km and $\pm$1500~km).  Apart from the run at the Kamioka site,
statistics are the same for all runs and error bars are shown as
examples on the points at $\pm2000$~km.  Statistics on the Kamioka
site points are a factor 10 larger.  Left panel: neutrinos in the
energy range 300 -- 500 MeV.  Right panel: neutrinos in the energy
range 1 -- 1.5 GeV.  }
\label{fig:kamns}
\end{figure}

Our 3D calculation uses a circular detector of radius 500 km.  We
study the flux obtained by positioning the detector in steps of 500 km
North, South, East and West from the real experimental location.  Note
that these displaced runs have a factor 10 lower Monte-Carlo
statistics than the main runs.  Figure~\ref{fig:kamns} shows the
zenith angle distribution of the flux of 300--500 MeV neutrinos for
various distances away from Kamioka in increments of 500km from 2000km
south to 2000km north.  The flux varies by a factor of 3 over the
range explored.  Furthermore, the shape of the zenith angle
distribution changes considerably over this range. Note the region in
the upward direction (cos($\theta_Z$) between -0.9 and -0.5) where the
variation of flux with latitude reverses compared to the rest of the
distribution.  Such variations are to be expected because the cutoffs
change considerably as a function of latitude.

As the neutrino energy increases, the variation decreases somewhat.
In the neutrino energy range 0.5~--~1~GeV, the variation is about a
factor 2.3 near the horizon.  The second panel in
figure~\ref{fig:kamns} shows the distributions for the neutrino energy
range 1 -- 1.5 GeV where the variation over the $\pm2000$~km range
studied is still about a factor of 2.  For neutrino energies above
2~GeV, the variation is considerably smaller -- about 20\%.

The variations in the east-west direction are far less.  This is
expected since we are moving in a direction where the geomagnetic
latitude  doesn't change so rapidly.  The fluxes vary by less than 10\% over
the 4000~km range studied in the 300~--~500~MeV neutrino energy range.
The flux is not entirely flat in this direction since the direction of
constant geomagnetic latitude is not exactly aligned with the
East-West direction and because the field is not exactly a dipole.

\begin{figure}[phtb]
\begin{center}
\epsfig{file=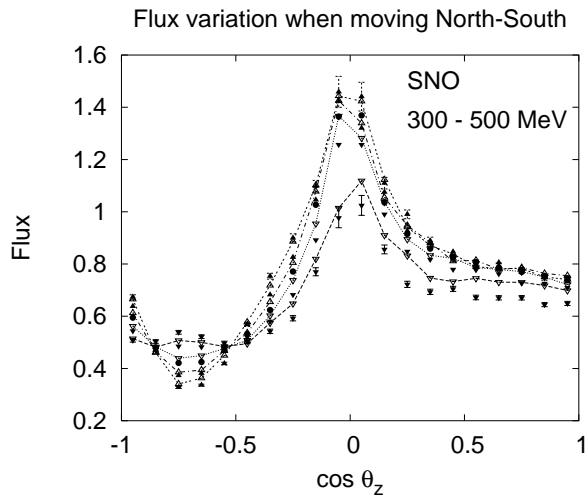,width=0.49\textwidth}
\end{center}
 \caption{Results of runs spaced 500 km apart in latitude for neutrinos 
in the energy range 300~--~500~MeV around the SNO location.  The legend 
is the same as in figure~\ref{fig:kamns}
}
\label{fig:snons}
\end{figure}

The zenith angle distribution at SNO is shown in
figure~\ref{fig:snons}; the variations are considerably smaller than
at Kamioka.  Tserkovnyak et al~\cite{Waltham} reported a `cliff'
effect in the geomagnetic cutoffs near the SNO site; this can be seen
in the downward fluxes on figure~\ref{fig:snons} -- moving north from
SNO causes a tiny ($<5$\%) variation, whereas moving south causes a
20\% variation in the fluxes.  The reason the effect on the fluxes is
considerably less at SNO than at Kamioka is because the cutoffs at
Kamioka are close to 20~GeV which is the most important primary energy
for the production of neutrinos of these low energies; any change in
the cutoffs has a dramatic effect on the fluxes.  At SNO, however, which is much
closer to the geomagnetic poles, the vertical cutoffs are around
2~GeV.  Since this is below the main primary energy range for neutrino
production, the effect as a function of latitude is smaller.  The
east-west variation around the SNO site is also minimal, in much the
same way as at Kamioka.
These conclusions are also valid for the Soudan detector site which is
only 1.4$^\circ$ (150~km) north and 11$^\circ$ (840~km) west of SNO.

 To determine what detector size is appropriate for use in a flux
 calculation, the functional form of the flux variation with both
 latitude and longitude is studied in each zenith angle bin separately.
 If the flux variation is linear within the range used for a simulated
 detector (and the simulated detector is centered on the real detector
 location), then the average flux across the simulated detector will be
 equal to the flux at the center of the detector.  Any non-linear
 variation however will result in a difference in fluxes.  If the flux
 is parameterized using a quadratic $\phi(x) = A_0 + A_1x + A_2x^2$,
 where $x$ is the distance as measured from the detector center, then the
 correction $C$ for using a large rectangular detector (from $x=-\Delta$ to $x=+\Delta$) in the simulation is
\begin{equation}
C = {{{1 \over 2\Delta}\int^{+\Delta}_{-\Delta} \phi(x) \mathrm{d}x - \phi(0) }\over {\phi(0)}} 
  = {{A_2\Delta^2} \over {3 A_0}}
\label{eqn:a2}
\end{equation}
 The corresponding expression for a circular (or elliptical) detector 
 shape of radius $r$ is $C=A_2r^2/4A_0$.

\begin{figure}[htb]
\begin{center}
\epsfig{file=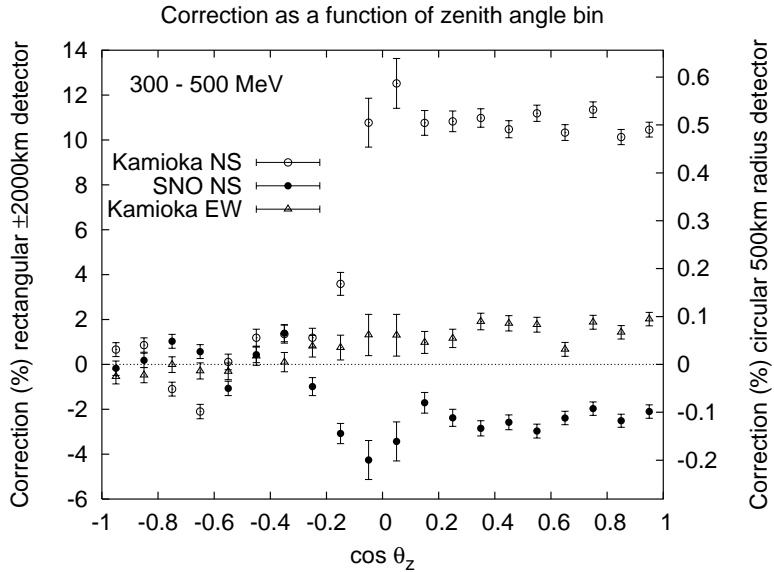,width=0.7\textwidth}
\end{center}
 \caption{Plot of the correction to be applied to the fluxes if a
large detector is used in the calculations.  The left scale gives 
the size of correction for a rectangular detector with extent 
$\pm$2000~km about the Kamioka site and the right scale gives the 
correction for a circular (or elliptical) detector with radius 500~km.}
\label{fig:Cplot}
\end{figure}
 
Quadratic fits to the variation in each zenith angle bin are made to
determine the values of $A_2/A_0$ for use in the expression given in
equation~\ref{eqn:a2}.  The resulting corrections $C$ are shown in
figure~\ref{fig:Cplot} for neutrino energies 300~--~500~MeV.  Since
the expressions for $C$ vary only by a scale factor proportional to
the detector size squared, these numbers can be scaled to any detector
size.  The North-South corrections at Kamioka are the largest; with a
$\Delta = 2000$~km detector $C$ is about 10\%.  With the detector size
used in this paper ($r=500$~km, shown on the right hand scale of
figure~\ref{fig:Cplot}) $C$ remains below 0.5\% for all zenith angle
bins.  Figure~\ref{fig:Cplot} also shows $C$ for the North-South
variation at SNO (of opposite sign to the Kamioka correction to
correct for the `cliff' effect) and the East-West correction at
Kamioka (which is considerably smaller than the North-South variation.
The East-West variation at SNO (not shown) is also negligible (i.e.
$C<0.5$\% for a 2000~km detector).  Similar curves have also been made
for higher energies.  The corrections show similar features to those
illustrated but with diminishing size as the energy increases.  The
average $C$ for downward fluxes at Kamioka (SNO) is
        5.1\% (-1.1\%) $\pm 0.3$\% in the energy range 0.5~--~1~GeV and 
        1.5\% (-0.2\%) $\pm 0.5$\% in the energy range 1~--~1.5~GeV.
At higher energies, $C$ becomes negligible.

In summary, the investigation of the variation of the calculated
neutrino fluxes with latitude has revealed large variations,
particularly at Kamioka where the variation in local cutoffs affects
the most important energy range of cosmic rays for low energy neutrino
production. These variations indicate that caution should be used in
deciding how much to extend the size of the detector in a 3D
calculation.  The present calculation uses a detector which is
circular of radius $r=500$~km which is sufficiently small to keep
corrections less than 0.5\%.  We elect to keep this detector size and not
to make any correction on the data -- possibly in the future, we will
use an elliptical detector with $r=500$~km in the North-South
direction and larger in the East-West direction.

Variations in the altitude of the detector have also been studied.  An
increase of the geometrical 3D enhancement near the horizon (5\% per
kilometre rise in altitude for 300 to 500~MeV neutrinos) is observed,
as expected~\cite{Lipari1} since the detector is moved closer towards
the production altitude.

\section{Conclusions}

We have extended the 1-dimensional neutrino flux
calculation~\cite{BGS,AGLS} to three dimensions.
We compare results of the 3D and 1D calculations for
neutrino energies from 100 MeV to 10 GeV over the full 
$4\pi$ solid angle.  Results are given for Kamioka,
which is at low geomagnetic latitude, and for SNO and Soudan
at high latitude.

The angle-averaged fluxes are identical within statistics
for $E_\nu>1$~GeV, with approximately a 3\% excess in the 3D
calculation for sub-GeV neutrinos.  The differences are much
more noticeable in the zenith angle distributions
for $E_\nu< 1$~GeV, which
show a significant excess in the 3D calculation 
for $-0.1<\cos\theta< 0.1$ and
a smaller deficit for $|\cos\theta|> 0.4$.  The differences
largely cancel in the angular integral, leaving the small (3\%)
overall difference mentioned above.  The zenith-angle differences
decrease with energy and become completely negligible for 
$E_\nu> 5$~GeV.  

Corresponding to the horizontal excess in the 3D calculation is
a contribution to the path length distribution of near
horizontal neutrinos which reduces the average path length
for horizontal neutrinos with $300< E_\nu< 500$~MeV
by 15\% relative to the 1D calculation.  The corresponding
reduction decreases to 4\% for $E_\nu > 1$~GeV.
This difference could have a small effect on the inferred
neutrino mass squared difference ($\Delta m^2$), but this 
remains to be determined.

The angle-integrated neutrino flavor ratio, calculated as 
$(\nu_\mu + {1\over 2}\overline{\nu}_\mu)\,/\,(\nu_e+{1\over 2}\overline{\nu}_e)$
is the same in the 3D as in the 1D calculation.  The
$\langle\nu\rangle\,/\,\langle\overline{\nu}\rangle$ ratio 
(where $\langle ...\rangle$
indicates angle averaging) is also the same.  The ratio of 
downward to upward neutrinos ($\cos\theta > 0.4$ and $\cos\theta< -0.4$)
are nearly identical in the 3D and 1D calculations.  However the 
ratio of downward to horizontal fluxes changes considerably below 
1~GeV when moving to a 3D calculation.
Because the neutrino flavor ratio and the up-down ratio are the most
important quantities for determining the neutrino oscillation
parameter $\sin\theta_{23}$, we do not expect correct treatment of the
3D effects in itself to lead to a large change in inferred value of
this parameter.  The oscillation parameter $\Delta m_{23}^2$ however is
also sensitive to the down-horizontal ratio and 3D effects could
possibly lead to a change in this parameter, particularly if low
energy horizontal neutrinos are used in the analysis.

The azimuthal distributions in the 3D calculation show interesting
effects which depend on neutrino flavor and also distinguish
between neutrinos and anti-neutrinos.  These differences arise
from bending of charged particles inside the atmosphere.  Unlike 
enhancement of low-energy neutrinos near the horizon (which is
a geometrical effect) this geomagnetic effect persists to higher
energy.  Because of its $\nu-\bar{\nu}$-dependence,
this generalized East-West effect could be detectable
in a sufficiently large detector with the capability
of determining the sign of neutrino-induced leptons.  Because
$\sigma_\nu\,>\,\sigma_{\bar{\nu}}$ for both neutrino flavors
while the relation between muon charge and the lepton number of
its neutrino decay product is opposite for $\nu_e$ and $\nu_\mu$, this
effect may also be visible as a difference in the East-West
effect for electron-like and muon-like events~\cite{Lipari2}.

We have shown that the treatment near the horizon in a 3D calculation
requires special consideration, in particular in handling the
weighting when a flat detector is used.  Computational speed is a major
concern in a 3-D calculation.  The detector may be made artificially
large to average the neutrino flux over a larger area around the
detector site.  We investigated of variation of the
flux with distance from the detector site and find that the changes
are dramatic.  However, the variation of the flux remains linear for a
significant distance and the use of a detector with an extent of the
order of 1000km may be used with care.  The problems are largest when
moving in a North-South direction at a detector located at low
geomagnetic latitudes (e.g. Kamioka).

In a forthcoming paper we will consider the 3D flux for different assumptions
about the primary spectrum.  We expect the limiting factors in the knowledge
of neutrino fluxes (at production, before
oscillations) to be uncertainties in the primary spectrum and in the
treatment of hadronic interactions, rather than 3D effects.

\subsection*{Acknowledgements}

The work of TKG and TS is supported in part by the U.S. Department 
of Energy under DE-FG02 91ER 40626.
One of the authors (SR) is supported by a PPARC studentship.

\end{document}